\documentclass{aa}  

\usepackage{graphicx}
\usepackage{txfonts}
\usepackage{upgreek}
\usepackage{enumitem}
\usepackage{xcolor}

\usepackage{supertabular}
\usepackage{booktabs}
\usepackage{multirow}

\usepackage{siunitx}
\DeclareSIUnit\mag{mag}
\DeclareSIUnit\pc{pc}
\DeclareSIUnit\mas{mas}
\DeclareSIUnit\au{au}
\DeclareSIUnit\msun{\ensuremath{M_\sun}}

\usepackage{array}
\newcolumntype{H}{>{\setbox0=\hbox\bgroup}c<{\egroup}@{}}

\newcommand*\mean[1]{\langle #1 \rangle}
\bibpunct{(}{)}{;}{a}{}{,} % to follow the A&A style

\begin{document} 
\graphicspath{{../images/}}   
   \title{Interferometric study on the temporal variability of the brightness distributions of protoplanetary disks}
   
   %\titlerunning{Tracing temporal variations in potential planet-forming regions}

   \author{     J. Kobus \inst{1}
                        \and
                S. Wolf \inst{1}
                                \and
                T. Ratzka \inst{2,3}
                                \and
                R. Brunngräber \inst{1}
          } 

   \institute{Institute of Theoretical Physics and Astrophysics, University of Kiel, Leibnizstraße 15, 24118 Kiel, Germany \and
  Institute for Physics/IGAM, NAWI Graz, University of Graz, Universitätsplatz 5/II, 8010 Graz, Austria \and
  Planetarium im Museum am Schölerberg, Klaus-Strick-Weg 10, 49082 Osnabrück, Germany \\
              \email{jkobus@astrophysik.uni-kiel.de}}

   \date{}

  \abstract
  % context heading (optional), leave it empty if necessary  
 {Optical and infrared spatially unresolved multi-epoch observations have revealed the variability of pre-main sequence stars and/or their environment. Moreover, structures in orbital motion around the central star, resulting from planet-disk interaction, are predicted to cause temporal variations in the brightness distributions of protoplanetary disks. Through repeated observations of pre-main sequence stars with the Very Large Telescope Interferometer (VLTI) over nearly two decades, the ESO Archive has become a treasure chest containing unprecedented high-resolution multi-epoch near- and mid-infrared observations of the potential planet-forming regions in protoplanetary disks. } 
  % aims heading (mandatory)
 {We aim to investigate whether the existing multi-epoch observations provide evidence for the variability of the brightness distributions of the innermost few astronomical units of protoplanetary disks and to quantify any variations detected.}
  % methods heading (mandatory)
 {We present different approaches to search for evidence of temporal variations based on multi-epoch observations obtained with the VLTI instruments PIONIER, AMBER, and MIDI for 68 pre-main sequence stars.}
  % results heading (mandatory)
  {For nine objects in our sample, multi-epoch data obtained using equal baselines are available that allow us to directly detect variations in the visibilities due to temporally variable brightness distributions. Significant variations of the near-infrared visibilities obtained in different epochs with PIONIER and/or AMBER for HD 50138, DX Cha, HD 142527, V856 Sco, HD 163296, and R CrA were found. HD 37806, TW Hya, and CPD-36 6759 show no significant variations. 
 By estimating the impact of a small variation of the baseline on the measured squared visibilities, we are able to compare the data of another 12 pre-main sequence stars. Thereby, we find evidence for temporal variations of the brightness distribution of one additional object, AK Sco. Besides the two binaries DX Cha and AK Sco, HD 50138 and V856 Sco also show signs of variability caused by variations of asymmetric structures in the brightness distribution.%The considered MIDI data does not allow us to derive any conclusions about the variability of the objects in the mid-infrared.
 }
  % conclusions heading (optional), leave it empty if necessary 
   {}


   \maketitle

%================================================================================
%================================================================================
\section{Introduction}\label{sec:introduction}
%================================================================================
Observations of pre-main sequence stars at different evolutionary stages and the subsequent investigation of the processes involved in the dispersal of their circumstellar disks, and hence also in the formation of planets, are the prerequisites for the development of planet formation models. Multi-epoch observations are particularly useful in this context, as they allow us to study the dynamics in the disks. 

In \citeyear{1945CMWCI.709....1J}, \citeauthor{1945CMWCI.709....1J} observed photometric variability of up to 3 mag for 11 T Tauri stars, of which five are also part of this study (R CrA, S CrA, RU Lup, T Tau, RY Tau). Over recent decades it has turned out that variability in the X-ray to mid-infrared wavelength range is common for pre-main sequence stars \citep[e.g.,][]{2020MNRAS.tmp..208S, 2019A&A...628A..74G, 2019ApJ...874..129R, 2015AJ....150..132R, 2015A&A...581A..66V, 2012ApJS..201...11K}. Recent infrared observations of several clusters with the Spitzer Space Telescope at $3.6$ and $\SI{4.5}{\micro\metre}$ revealed that typically more than half of the pre-main sequence stars show significant temporal variations \citep[e.g., Serpens South: $\SI{\sim70}{\percent}$, NGC 1333: $\SI{\sim50}{\percent}$, GGD 12-15: $\SI{\sim85}{\percent}$, CSI 2264: $\SI{\sim90}{\percent}$;][]{2018AJ....155...99W, 2015AJ....150..175R, 2015AJ....150..145W, 2014AJ....147...82C}. 

Different mechanisms lead to such variations: \textsf{Type I}) Cool spots on the stellar surface resulting from the interaction of the stellar magnetic field with the photospheric gas are stable over many rotational periods of the star. Due to the asymmetric distribution, with coverage of a few percent of the stellar surface, and temperature contrasts of $\SI{\sim1000}{\kelvin}$, the stellar rotation induces periodic variations with periods of several hours to a few days \citep{1994AJ....108.1906H}. At $3.6$ and \SI{4.5}{\micro\metre}, the variability due to cool spots is typically about 0.08\,mag \citep{2015AJ....150..118P}. Additional long-term variations of the spot properties on timescales of months or years lead to changes in the shape and amplitudes of the light curves \citep{2008A&A...479..827G}. \textsf{Type II}) Hot spots result from the magnetically channeled accretion of circumstellar material onto the star. Similar to their cool counterparts, they induce photometric variability due to the stellar rotation. However, variations in the accretion with timescales of hours to years \citep{2016ARA&A..54..135H} induce additional irregular variability \citep{1994AJ....108.1906H}. Typical amplitudes in the mid-infrared are $\SI{\sim0.15}{mag}$ \citep{2015AJ....150..145W, 2015AJ....150..118P, 2012ApJS..201...11K}. \textsf{Type III}) Variations of so-called 'dippers' showing day-long dimmings with depths of up to several tens of percent are thought to be caused by circumstellar material obscuring the stellar emission. The cause of these occultations is still an open question. One possibility are asymmetric variations in the vertical dust density distribution, such as spiral arms, of nearly edge-on disks. Recent studies, however, indicate that misaligned inner disks are common \citep{2020MNRAS.492..572A}, also explaining dipper events for pre-main sequence stars that have an outer disk with a lower inclination. Depending on the radial distance of the light attenuating material from the star, the timescale for these variations ranges from days to years.

Circumstellar material influences the observed temporal variations in various ways. Additionally to a variable accretion onto the star or obscuration, the emission of circumstellar dust can directly contribute to photometric variations in the infrared. A significant amount of the infrared radiation is emitted from the hot and warm dust in the innermost region of the circumstellar disk. A variable accretion rate, and thus variable luminosity of the central source, will result in a changing illumination as well as in fluctuations of the temperature distribution of the dust in the inner disk, to eventually become observable as a temporal variation of the infrared brightness distribution \citep{2001AJ....121.3160C}.

While the identification of the type of variability is already possible based on unresolved multi-wavelength observations with sufficient temporal sampling, high-resolution observations offer the opportunity to investigate the impact of the circumstellar disk structure on the photometric variabilities observed. Furthermore, such observations enable us to study another intriguing type of temporal variability: variations in the brightness distribution due to the orbital motion of asymmetric features in the disk. These asymmetries, such as spiral density arms and local hot spots, may result from planet-disk interactions, allowing us to derive planetary and disk parameters \citep{2018A&A...611A..90B, 2014A&A...572L...2R, 2013A&A...549A..97R, 2005ApJ...619.1114W}.

In this context, long-baseline interferometry with the Very Large Telescope Interferometer (VLTI) provides exceptional opportunities. Covering the near- and mid-infrared atmospheric windows, the VLTI is sensitive to scattered stellar radiation as well as to thermal emission from hot and warm dust in the innermost potential planet-forming region of protoplanetary disks. With spatial resolutions down to a few milliarcseconds (mas), it thereby allows us to resolve sub-au scale structures of protoplanetary disks in nearby star-forming regions. The potential of multi-epoch VLTI observations for studying the variability of the brightness distribution has already been demonstrated in a few studies. Comparing mid-infrared observations obtained with the VLTI instrument MIDI \citep[MID-infrared Interferometric instrument,][]{2003Msngr.112...13L}, \citet{2016A&A...585A.100B} found a significant change in the radial brightness distribution of DR Tau suggesting that
the hot inner region of the disk appeared more compact in January 2005 than in December 2013. Near-infrared observations with the VLTI instrument PIONIER \citep[Precision Integrated-Optics Near-infrared Imaging ExpeRiment,][]{2011A&A...535A..67L} enabled the motion of a brightness asymmetry in the disk around MWC 158 to be traced \citep[HD 50138, ][]{2016A&A...591A..82K}.

Since the start of scientific observations with the VLTI almost two decades ago, a significant amount of multi-epoch archival data has become available, potentially providing an unprecedented basis for studying the variability of the brightness distribution of dozens of pre-main sequence stars. We aim to investigate to what extent the existing multi-epoch observations allow us to derive conclusions about the variability of the brightness distributions of pre-main sequence stars and the innermost region of their protoplanetary disks. Furthermore, it is our goal to quantify the variations found. In Sect. \ref{sec:obsAndDataRed}, we present the sample containing 68 pre-main sequence stars and the near- and mid-infrared observations obtained with the instruments PIONIER, AMBER \citep[Astronomical Multi-BEam combineR,][]{petrov_amber_2007}, and MIDI. We introduce our procedure for the analysis of the interferometric multi-epoch observations in Sect. \ref{sec:procedure}. In Sect. \ref{sec:results}, we present the results on the temporal variability based on the different methods. We discuss previous variability studies and the reasons for unidentified variations and finally classify the detected variations in Sect. \ref{sec:disc}. We conclude our findings on the variability of the near- and mid-infrared brightness distributions in Sect. \ref{sec:concl}.

%================================================================================
%================================================================================
\section{Observations and data reduction\label{sec:obsAndDataRed}}
%================================================================================
\subsection{Sample description}
This study is based on archival near- and mid-infrared interferometric data obtained with VLTI instruments PIONIER, AMBER, and MIDI. We used the VLTI/MIDI atlas \citep{2018A&A...617A..83V} containing 82 low and intermediate mass young stellar objects as a basis for our target selection. For these 82 objects, we searched for archival multi-epoch observations covering at least two nights with a minimum time lag of one day, allowing us to investigate variability on different temporal scales from short term (one to a few days; e.g., tracing outbursts) to long term (months to years; e.g., tracing the orbital motion of disk asymmetries). In total, we found 68 objects; of these, multi-epoch observations of 35 are available with PIONIER, 16 with AMBER, and 49 with MIDI. An overview of the sample is given in \mbox{Table \ref{tab:sample}}.

Eight objects in our sample (T Tau S, GG Tau, GW Ori, Z CMa, DX Cha, HD 142527, AK Sco, and R CrA) are close binaries with separations smaller than the field of view \citep[\SI{\sim 0.16}{\arcsec},][]{2016SPIE.9907E..3BH}. For these we expect temporal variations of the brightness distribution due to the orbital motion of the stellar companions. Identifying these variations is irrelevant in terms of studying the different causes of variability discussed in the introduction. However, these objects offer the possibility to get an impression of how effectively the available data and different approaches enable the identification of temporal variability. 

For 8 out of the 35 objects observed with PIONIER, observations from only 2 nights each are available (CO Ori, GW Ori, V1247 Ori, FU Ori, TW Hya, CV Cha, HD 141569, and SR 24A). HD 142527 was most frequently observed with PIONIER with 22 nights. On average, the 68 objects were observed in 6 nights with PIONIER; the median is 4 nights. The timescales covered by the PIONIER observations range from one day (UX Ori, CQ Tau, HD 37806, HD 50138, DI Cha, WW Cha, DX Cha, CPD-36 6759, HD 142527, RU Lup, HD 144432, V856 Sco, and HD 163296) to 2579 days (seven years, FU Ori). The mean time lag between two observations is 161 days. The median time lag amounts to 18 days.

In the case of AMBER, 3 objects have been observed only in 2 nights each (RY Tau, HD 45677, and HD 142666). With observations from 33 nights, V856 Sco was most frequently observed with AMBER. The mean value of the number of nights for which AMBER observations are available per object is 9. The median is 8 nights. The timescales covered by the AMBER observations range from one day (HD 259431, HD 50138, HD 85567, HD 100453, DX Cha, CPD-36 6759, HD 142527, V856 Sco, HD 150193, R CrA) to 2905 days (8 years, HD 50138). The mean time lag between two observations is 176 days with a median of 23 days.

MIDI observations on only 2 nights each have been performed for 8 objects (HD 31648, Ass Cha T 1-23, CV Cha, HD 141569, GSS 31, SR 21A, AS 209, VV CrA NE). Z CMa was most frequently observed with MIDI on 20 nights. The mean value of the number of nights per object is 5, the median is 4 nights. The MIDI observations cover timescales from one day (DG Tau, GG Tau, AB Aur, GW Ori, HD 36112, Z CMa, HD 72106, TW Hya, DI Cha, HP Cha, WW Cha, DX Cha, HD 142527, HD 325367, HD 144432, V856 Sco, GSS 31, Elia 2-24, Haro 1-16, V346 Nor, HD 150193, HD 163296, HD 169142, S CrA N, VV CrA SW, VV CrA NE, and HD 179218) to 3764 nights (10 years, V856 Sco). Here, the mean value of the time lag between two observations is 338 days, with a median of 32 days.

\subsection{Observations}
The instrument PIONIER combines either the four 8.2-m Unit Telescopes (UTs) or the four 1.8-m Auxiliary Telescopes (ATs) of the VLTI, leading to simultaneous visibility measurements at six baselines together with four closure phases. Observations can be performed with one (spectral resolution $R \approx 5$) to six ($R \approx 30$) spectral channels, and in early observations up to seven spectral channels over the $H$ band (central wavelength $1.65\,\upmu$m). A spatial resolution down to about \SI{2}{mas} ($\approx\!0.1\,$au for the nearest object TW Hya at 60\,pc) is reached for the longest baseline available ($\sim\!130\,$m).
The observations with PIONIER included in this study have been performed on 105 nights between December 2010 and June 2018. A summary of the considered PIONIER observations is given in Table \ref{tab:logPIO}.

AMBER was a three-beam combining instrument for the $H$ and $K$ (central wavelength $2.2\,\upmu$m) bands, providing visibility measurements at three baselines together with one closure phase per observation (decommissioned in 2018). Different spectral settings were offered allowing observations at low ($R\approx35$) and medium ($R\approx1500$) spectral resolution in both the $H$ and $K$ bands, as well as high resolution observations ($R\approx12000$) in the $K$ band. The AMBER data considered in this study were obtained on 123 nights between May 2007 and December 2015. A summary of the considered AMBER observations is given in Table \ref{tab:logAMB}.

The instrument MIDI was a mid-infrared ($8 - 13\,\upmu$m) beam combiner operating between 2003 and 2015. Combining the signal from either two ATs or two UTs, MIDI delivered one visibility measurement per observation and no closure phase. Depending on the spectral dispersing element, MIDI allowed observations with spectral resolutions of $R\approx30$ (prism) and $R\approx260$ (grism). Due to the longer wavelength range targeted, MIDI had a lower spatial resolution of down to about \SI{10}{mas} corresponding to $\sim 0.6\,$au for TW Hya (at 60\,pc). The MIDI data contain observations from 201 nights between February 2004 and March 2015. A summary of the MIDI observations is given in Table \ref{tab:logMIDI}.

\subsection{Data reduction}
We took the already reduced and calibrated PIONIER data from the Jean-Marie Mariotti Center \texttt{OiDB} service\footnote{\url{http://oidb.jmmc.fr/index.html}}. The data reduction is based on \texttt{pndrs} package versions 2.3 to 3.74 \citep{2011A&A...535A..67L}. 

The AMBER data reduction was performed with the \texttt{amdlib} package version 3.0.9 using the procedure described in \citet{tatulli_amber_2007} and \citet{2009A&A...502..705C}. In order to avoid the attenuation of the fringe contrast induced by vibrations of the VLTI instrumentation, we only used 20\,\% of the frames with the highest signal-to-noise ratio \citep[$\mathrm{S/N}$,][]{tatulli_amber_2007}. Data from the $H$ or $K$ bands containing frames with an $\mathrm{S/N}\leq 1$ or a mean $\mean{\mathrm{S/N}}\leq 3$ after this selection were excluded bandwise. To further increase the data quality, we appended all observations that do not exceed a period of 25 minutes. For the calibration we used all available data obtained for suitable calibrator stars in the respective nights, which we downloaded from the ESO science archive\footnote{\url{http://archive.eso.org/eso/eso_archive_main.html}} together with the interferometric raw data of the science targets. The calibrator diameters used to calculate the transfer functions were collected from the calibrator catalogs by \citet{bourges_jmmc_2014}, \citet{lafrasse_building_2010}, \citet{richichi_first_2005}, \citet{merand_catalog_2005}, and \citet{borde_catalogue_2002}. Eventually, we excluded all observational data of the science targets for which the transfer function shows significant temporal variations.

Already reduced and calibrated MIDI data were taken from the VLTI/MIDI atlas\footnote{Available at \url{https://konkoly.hu/MIDI_atlas/}.} \citep[][]{2018A&A...617A..83V}. A detailed description of the data reduction procedure is given in \citet[][]{2018A&A...617A..83V}. We excluded data marked as unreliable during their reduction process.

%================================================================================
%================================================================================
\section{Procedure}\label{sec:procedure}
%================================================================================
\subsection{Comparing multi-epoch interferometric data: Approaches}\label{sec:approaches}
Our goal is to exploit the extensive set of archival near- and mid-infrared interferometric data covering almost two decades to investigate the variability of pre-main sequence stars.
The fact that one does not directly measure the brightness distribution of the object with an interferometer, but the Fourier transform for a limited set of spatial frequencies, has a decisive influence on the approach and possible constraints that can be derived in this investigation. The measured visibilities depend not only on the brightness distribution, but also on the projected\footnote{In the following we refrain from the adjective "projected", however, whenever mentioning the baseline, its length, or orientation, the projected baseline is referred to.} baseline, characterized by its length and orientation, and spectral wavelength range (band). Comparing interferometric measurements from different nights to study the variability of the brightness distribution directly thus requires the observations to be obtained for the same instrumental setup (i.e., using the same baseline and band).

Therefore, in the first part of this study, we compare multi-epoch observations that have been obtained using equal baselines and the same band. In this case, differences in the measured visibilities can be directly attributed to temporal variations in the brightness distribution. A detailed description of this approach and results on the temporal variability of objects that can be investigated based on this approach can be found in Sect.\ \ref{sec:dircomp}.

Multi-epoch observations using equal baselines, which are required for constraining temporal variability based on the direct comparison of visibilities, are only available for a small subset of the sample. Therefore, we present additional approaches to investigate temporal variability based on visibility data that have been obtained for deviating baselines. Here, the basic idea is to estimate the differences in the visibilities we expect due to small deviations of the baseline length and orientation and to compare these to the measured differences.

We elaborate on three different approaches to identify variability this way, focusing on the investigation of squared visibilities. In Sect. \ref{sec:Plot}, we present a qualitative approach for which we plot the measured visibilities against the baseline length and additionally indicate the baseline orientation and date. In Sect. \ref{sec:extComparison}, we estimate the typical visibility differences due to small changes of the baseline and compare these to the visibility differences we obtain from comparing data with moderately deviating baselines from different epochs. A further approach is based on the modeling of interferometric quantities with Gaussian brightness distributions and the subsequent comparison of these models found for different epochs. As we find that this approach provides false positive results, we present it together with a discussion of the reason for its inappropriateness in Appendix \ref{sec:fitting}. 

For each of these three approaches, we first present the basic idea and describe the practical implementation. Then, we evaluate the reliability of each approach by applying it to synthetic data with known brightness distributions. Therefore, on the one hand, we consider synthetic data based on models with variable brightness distributions to test whether the respective approach is in principle capable of identifying variability. On the other hand, we also apply the approaches to synthetic data with static brightness distributions in order to uncover pitfalls that may lead to a misinterpretation of the data, that is, to unjustified conclusions about their temporal variability ("false-positive"). Finally, we use those approaches found to provide reliable conclusions to constrain the temporal variability of the pre-main sequence stars in our sample.

\subsection{Synthetic observations\label{sec:synthObs}}
As outlined above, we use synthetic observations with known static and variable brightness distributions to evaluate the reliability of the different approaches used to constrain variability based on data obtained for deviating baselines.
We derive time-dependent brightness distributions from radiative transfer (RT) simulations, for a model of an exemplary pre-main sequence star surrounded by a protoplanetary disk with typical properties.
 
\paragraph{Model} The model consists of a pre-main sequence star with an effective temperature of $T_\star = 9750$\,K and a luminosity of $L_\star = 18\,\mathrm{L}_\odot$. These values were chosen in accordance with typical properties of intermediate mass pre-main-sequence stars \citep[Herbig Ae/Be stars; ][]{1998ARA&A..36..233W, 1960ApJS....4..337H} and, for instance, are similar to those of HD 141569 \citep{2015MNRAS.453..976F}. The central star is surrounded by a protoplanetary disk for which we choose a typical inner radius of $0.3\,$au \citep[e.g., ][]{2019PASP..131f4301W}. The dust density distribution in cylindrical coordinates $r,z$ is given by
\begin{equation}
\label{eq:density}
\rho (r, z) = \frac{\Sigma(r)}{\sqrt{2 \pi}\,h(r)} \exp{\left[-\frac{1}{2}\left(\frac{z}{h(r)}\right)^2\right]}, 
\end{equation}
where $h$ is the scale height
\begin{equation}
h(r)=h_{100} \left(\frac{r}{100\,\mathrm{au}}\right)^\beta
\end{equation}
with a reference scale height of $h_{100} = 15\,$au and a flaring parameter of $\beta = 1.2$ \citep[e.g., ][]{2019PASP..131f4301W}.
Following the work of \citet{1974MNRAS.168..603L} and \citet{0004-637X-495-1-385}, we use a surface density distribution
\begin{equation}
\label{eq:surfaceDensity}
\Sigma(r) =\Sigma_0 \left(\frac{r}{100\,\mathrm{au}}\right)^{-p} \exp{\left[-\left(\frac{r}{100\,\mathrm{au}}\right)^{2-p}\right]}
\end{equation}
with $p = 0.9$ \citep[e.g., ][]{2010ApJ...723.1241A}. The reference surface density $\Sigma_0$ is chosen to scale the dust disk mass to $10^{-4}\,$M$_\odot$ \citep[e.g., ][]{2019PASP..131f4301W, 2010ApJ...723.1241A}.

We assume the dust to consist of $62.5\,\%$ silicate and $37.5\,\%$ graphite \citep{1984ApJ...285...89D} with optical properties from \citet{1984ApJ...285...89D}, \citet{1993ApJ...402..441L}, and \citet{2001ApJ...548..296W}. The grain sizes range from $5\,\mathrm{}$ to $250\,\mathrm{nm}$ with a size distribution given by $n(a)\propto a^{-3.5} \mathrm{d}a$ \citep{1977ApJ...217..425M}.

\paragraph{Radiative transfer} The simulations are performed with the 3D Monte-Carlo continuum and line RT code \texttt{Mol3D} \citep{2015A&A...579A.105O}. Assuming a distance of 140\,pc and a disk inclination of \SI{30}{\degree} from face-on, we calculate ideal intensity maps for two wavelength intervals $1.5 - 2.4\,\upmu$m and $7.9 - 12.6\,\upmu$m, each containing 12 logarithmically distributed wavelengths. 

\paragraph{Temporal variations} 
Visibilities depend not only on the observed, potentially temporal variable brightness distribution, but also on the chosen baseline. The different combinations of a small or large impact of a baseline deviation on the visibilities and no or present temporal variations of the brightness distribution should therefore be considered by the different models used to simulate synthetic observations. For this purpose, we utilize two topologically different models, each with a static and a time-dependent version. By adjusting the brightness distribution from the radiative transfer simulation (reference), we obtain the following four different models.

\begin{figure}
\includegraphics[width=\linewidth]{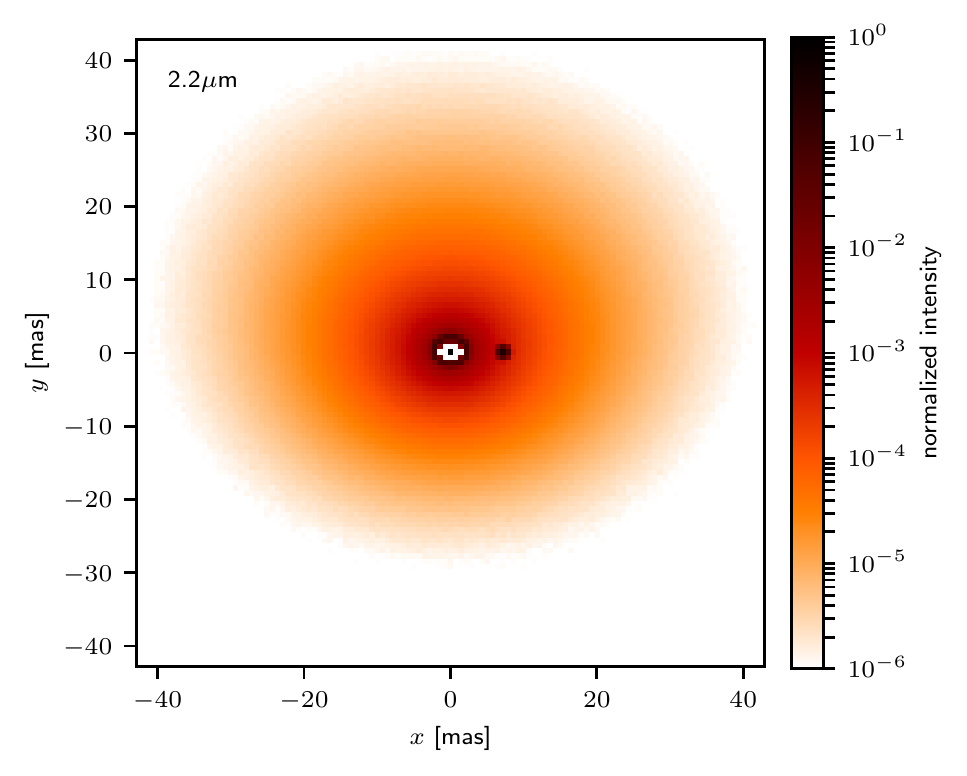} % [width=\linewidth]
\caption{Normalized intensity map of the model $M_{2,\mathrm{stat}}$ in the $K$ band (2.2$\upmu$m). This brightness distribution, which is based on radiative transfer simulations and supplemented by a Gaussian disk feature at \SI{9}{mas}, serves as a basis for synthetic observations, which are used to test the validity of different analysis approaches (see Sect.\ \ref{sec:procedure} for details).}
\label{fig:mapM2t2}
\end{figure}

\begin{itemize}[align=left]
\item[$M_{1,\mathrm{stat}}$:] The brightness distribution of this model corresponds exactly to the reference brightness distribution we obtain from the radiative transfer simulation. This model represents a time-independent brightness distribution with only a slight azimuthal asymmetry caused by the inclination of $30^\circ$ of the disk.

\item[$M_\mathrm{1}(t)$:] With this model we mimic temporal variations of the star-to-disk flux ratio as a simple representation of variability caused, for example, by stellar spots. In the reference brightness distribution, we vary the stellar intensity by multiplying it by a time-dependent factor that randomly switches between \num{0.5} and \num{1}. Therefore, we neglect the varying illumination of the circumstellar material and the fluctuation of the temperature distribution due to changes in the heating, which both can result in a temporal variation of the brightness distribution of the disk. The variation of the stellar intensity leads to a change of the total brightness of \SI{0.19}{mag} in the $K$ band, which corresponds to the observed variations caused by stellar spots \citep[e.g.,][]{2013ApJ...773..145W}.

\item[$M_{2,\mathrm{stat}}$:] For a simple model containing a significant asymmetry, we add a Gaussian feature to the reference brightness distribution. The Gaussian feature is positioned at a distance of \SI{9}{mas} (\SI{1.26}{au} at \SI{140}{pc}) to the central star with a standard deviation of \SI{0.5}{mas}. The intensity at the position of the Gaussian feature is increased by a factor of $10^3$ with respect to the local disk brightness we obtain from the radiative transfer simulation at this position, which is comparable to the relative intensity of embedded companions in the $N$ band simulated by \citet{2018A&A...611A..90B}. The resulting ratio of Gaussian feature to stellar flux is 0.11, 0.36, and 151 in the $H$, $K$, and $N$ bands, respectively. To illustrate this model, we show the brightness distribution at \SI{2.2}{\micro\meter} in Fig.\ \ref{fig:mapM2t2}.

\item[$M_\mathrm{2}(t)$:] For the time-dependent counterpart of model $M_{2,\mathrm{stat}}$, we vary the position of the Gaussian feature. Assuming Keplerian motion of the disk and a mass of the central Herbig star of \SI{1.8}{\ensuremath{M_\sun}} \citep{2015MNRAS.453..976F}, we adopt an orbital period of 400 days. The orbital distance is thereby varied according to the assumed inclination of \SI{30}{\degree}. The model is a simple representation of temporal variations due to the motion of disk asymmetries induced by planet-disk interactions.

\end{itemize}

\paragraph{Synthetic interferometric observations} We calculate synthetic visibilities and phases by performing a fast Fourier transform of the ideal intensity maps.
Therefore, both the UV coverage as well as the time lags between the observations are chosen to match the real PIONIER, AMBER, and MIDI observations of the objects in our sample (i.e., based on the four models, we calculate $140$, $64$, and $196$ synthetic multi-epoch PIONIER, AMBER, and MIDI observations, respectively). Finally, we add Gaussian noise with a standard deviation equal to the measured uncertainties to the synthetic visibilities and phases.

These synthetic data sets are not intended to provide the basis for modeling the underlying physics of the variability of the objects in our sample. They rather mimic visibilities comparable to the visibilities measured for the objects in our sample with the time-dependent synthetic data, thus providing significant variations comparable to the visibility differences found for objects showing evidence for temporal variability (see Sect.\ \ref{sec:results}). Mimicking the qualitative behavior of two fundamentally different types of variability, namely symmetrical variations, for example caused by the rotation of a star with cold spots, and the motion of a disk asymmetry, for example caused by an accreting protoplanet embedded in the disk, the synthetic data based on the time-dependent models $M_\mathrm{1}(t)$ and $M_\mathrm{2}(t)$ provide an opportunity to evaluate the ability of different approaches presented in Sect.\ \ref{sec:results} to identify variability. The time-independent synthetic data based on the models $M_{1,\mathrm{stat}}$ and $M_{2,\mathrm{stat}}$ enable us to evaluate the reliability of the approaches by investigating whether their application leads to false-positive results.

%================================================================================
%================================================================================
\section{Results \label{sec:results}}
%================================================================================
In the following sections, we investigate whether the available multi-epoch observations indicate temporal variations in the brightness distributions of the 68 pre-main sequence stars in our sample. For this purpose, we first analyze data for which observations using equal baselines have been performed in different epochs in Sect.\ \ref{sec:dircomp}. The direct comparison of visibilities measured at different times using equal baselines allows us to directly constrain temporal variability best.
However, for most of the objects in our sample the multi-epoch observations have only been obtained for different baselines. Therefore, in Sects.\ \ref{sec:Plot} and\ \ref{sec:extComparison} (and Sect.\ \ref{sec:fitting}), we apply the other approaches outlined in Sect.\ \ref{sec:approaches}. 

\subsection{Direct comparison\label{sec:dircomp}}
In the available data we search for multi-epoch observations where the UV coverage allows a direct comparison of visibilities measured in different epochs. Due to the rotation of the earth, the baseline changes during the execution of the observation. In the considered PIONIER data the mean change of the baseline length and the position angle during the observations is $\mean{\Delta BL} = 0.1\,\%$ and $\mean{\Delta PA} = 0.2^\circ$, respectively. Due to the accumulation of data obtained within a period of 25 minutes, in the case of AMBER the baseline length and position angle vary by $\mean{\Delta BL} = 1.4\,\%$ and $\mean{\Delta PA} = 1.7^\circ$, respectively. In the reduced MIDI data, no information about the baseline deviation is available, however, since the typical execution time of an observing block with MIDI was 25 minutes, we assume the mean baseline variations to be comparable to those of the AMBER measurements. For consistency, we stick to the lowest values found for the PIONIER observations and consider baselines that differ by less than $0.1\%$ in length and $0.2^\circ$ in orientation to be "equal".

Applying this criterion, we find data allowing the intended analysis based on a direct comparison for nine of the objects. Seven objects (HD 37806, HD 50138, TW Hya, DX Cha, CPD-36 6759, HD 142527, and V856 Sco) have at least one pair of visibility measurements with equal baselines obtained with PIONIER, and five objects (DX Cha, HD 142527, V856 Sco, HD 163296, and R CrA) in the case of AMBER. The MIDI data set does not contain observations obtained for equal baselines for any of the objects.

Besides the equivalence of the baseline, the spectral coverage of the observations must also be comparable. Therefore, we compare only observations made with the same instrument in the same band. However, these may differ in their spectral resolution.
In order to enable a comparison in such cases,  we average squared visibilities obtained for $n$ wavelengths within a given band, weighted with their uncertainties $\sigma_i$:
\begin{equation}\label{eq:ave}
\mean{V^2} = \frac{\sum_{i=1}^{n} \frac{V^2\left(\lambda_i\right)}{\sigma_i^2}}{\sum_{i=1}^{n} \frac{1}{\sigma_i^2}}.
\end{equation}
The corresponding averaged uncertainty is
\begin{equation}\label{eq:ave_err}
\sigma_{\mean{V^2}} = \sqrt{\frac{1}{\sum_{i=1}^{n} \frac{1}{\sigma_i^2}}}.
\end{equation}
We then calculate the absolute difference of two mean squared visibilities obtained in epochs $j$ and $k$
\begin{equation}\label{eq:ave_diff}
\Delta \mean{V^2} = |\mean{V_k^2} - \mean{V_j^2}|,
\end{equation}
together with its total uncertainty
\begin{equation}\label{eq:sigma}
\sigma = \sqrt{\sigma_{\mean{V_j^2}}^2 + \sigma_{\mean{V_k^2}}^2}.
\end{equation}
By comparing the difference of the squared visibilities to the total uncertainty $\Delta \mean{V^2} / \sigma$ we can calculate the significance of any variations detected.

\begin{figure*}
\includegraphics[]{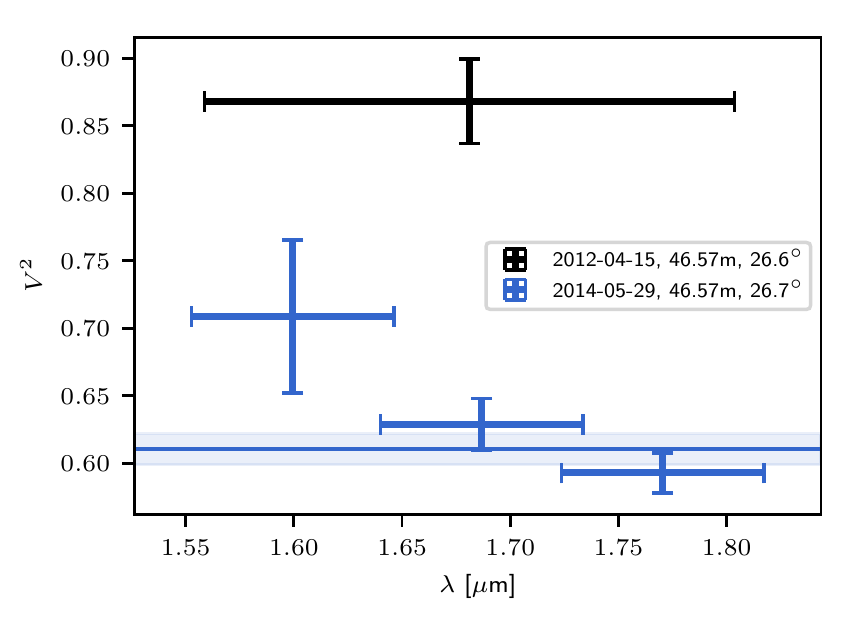}
\includegraphics[]{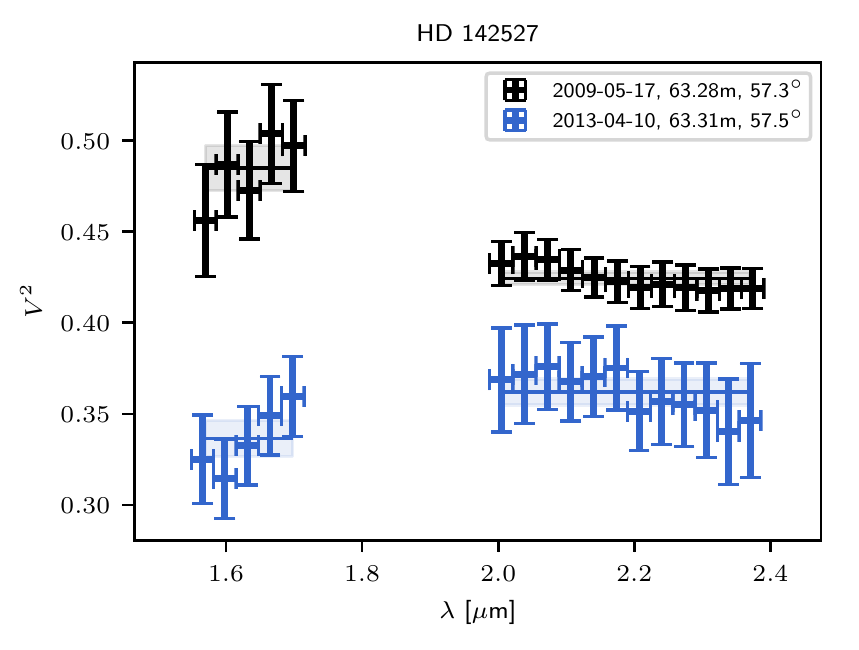}
\caption{\label{fig:comBLex} Squared visibilities obtained for HD 142527. Left: Squared visibilities obtained with PIONIER on two nights using equal baselines. Right: Squared visibilities obtained with AMBER on two other nights using equal baselines. The error bars in vertical direction give the uncertainties of the squared visibilities. In a horizontal direction, the bars indicate the spectral bins. Spectrally resolved measurements are averaged according to Eq.\ \ref{eq:ave}, where the mean values are plotted as solid lines and the corresponding uncertainties (Eq.\ \ref{eq:ave_err}) as shaded rectangles.}
\end{figure*}

On the basis of one example showing evidence for temporal variability, we demonstrate how we compare the visibilities obtained in different epochs using equal baselines and how we subsequently decide about the variability of the brightness distribution. For this purpose, we plot one pair of squared visibilities obtained for \mbox{HD 142527}  using equal baselines in two epochs  in Fig.\ \ref{fig:comBLex} (left: PIONIER, right: AMBER). In the left panel of this figure one can see that the $H$ band squared visibilities obtained with PIONIER on 29 May \mbox{2014} ($BL = 46.6\,$m, $PA = 26.6^\circ$) are significantly below those measured two years earlier (on 15 April 2012, $BL = 46.6\,$m, $PA = 26.7^\circ$). The difference in the wavelength-averaged squared visibilities amounts to $\Delta\mean{V^2}  = 0.26 \hat{=} 7.7\sigma$ (see Eqs.\ \ref{eq:ave_diff} and \ref{eq:sigma}). The squared visibilities obtained with AMBER (right panel, $BL \approx 63\,$m, $PA \approx 57^\circ$) for a different spatial frequency also decreased significantly between 17 May 2009 and 10 April 2013, both in the $H$ and $K$ bands. The squared visibility differences for the AMBER observations are $0.15 \hat{=} 9.4 \sigma$ and $0.06 \hat{=} 8 \sigma$ in the $H$ and $K$ band, respectively. Thus, both $H$ and $K$ band observations obtained with PIONIER and AMBER at different spatial scales indicate a significant change in the brightness distribution of HD 142527. 

\begin{figure*}
\includegraphics[width=\linewidth]{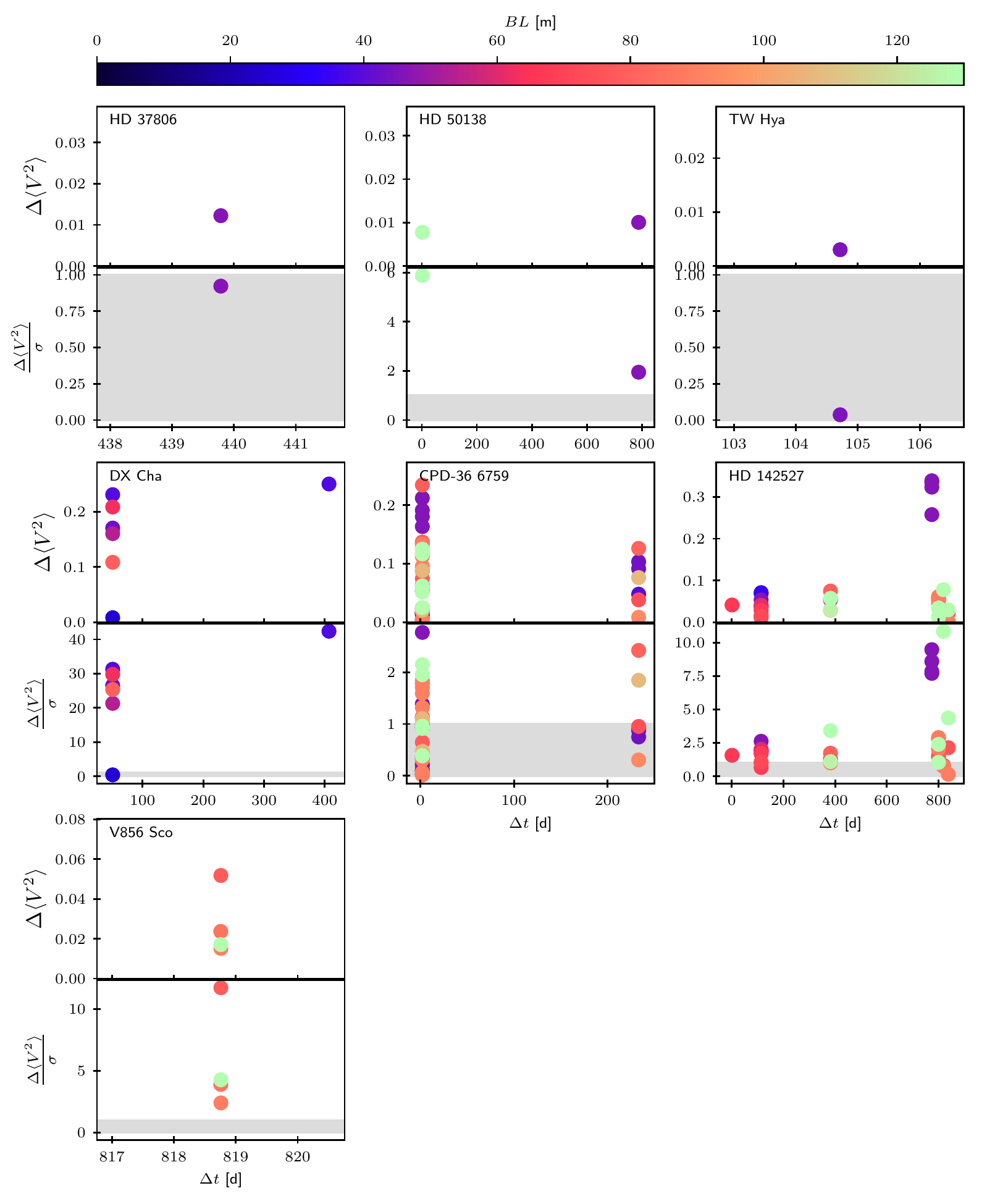}
\caption{\label{fig:sigmaPIONIER} 
Direct comparison of multi-epoch PIONIER observations obtained using equal baselines for HD 37806, HD 50138, TW Hya, DX Cha, CPD-36 6759, HD 142527, and V856 Sco. Upper panels: Absolute differences of the wavelength averaged squared visibilities (Eq.\ \ref{eq:ave_diff}) of all pairs obtained using equal baselines plotted against the time lag $\Delta t$. The baseline is indicated by the color. Lower panels: Absolute differences of the wavelength averaged squared visibilities divided by the uncertainty $\sigma$ (Eq.\ \ref{eq:sigma}). The gray shaded regions indicate where the differences are equal within the error bars ($\Delta \mean{V^2} / \sigma \leq 1$, see Sect.\ \ref{sec:dircomp} for details).
}
\end{figure*}

In total, 344 squared visibilities have been obtained for HD 142527 between March 2012 and June 2014 in the considered PIONIER observations. These can be combined to \num{58996} pairs, of which the squared visibilities of \num{56700} pairs were obtained in two different epochs (i.e., they belong to different observing blocks). Out of these, the squared visibilities have been obtained for 34 pairs using equal baselines ($\Delta BL < 0.1\%$, $\Delta PA < 0.2^\circ$). For these pairs a direct comparison of the squared visibilities allows us to directly constrain temporal variations. For this purpose, we calculate the differences in the wavelength-averaged squared visibilities (Eq.\ \ref{eq:ave_diff}) and the corresponding uncertainty $\sigma$ (Eq.\ \ref{eq:sigma}) for these 34 pairs (see middle plot in lower row of Fig.\ \ref{fig:sigmaPIONIER}).

\begin{figure*}
\includegraphics[width=\linewidth]{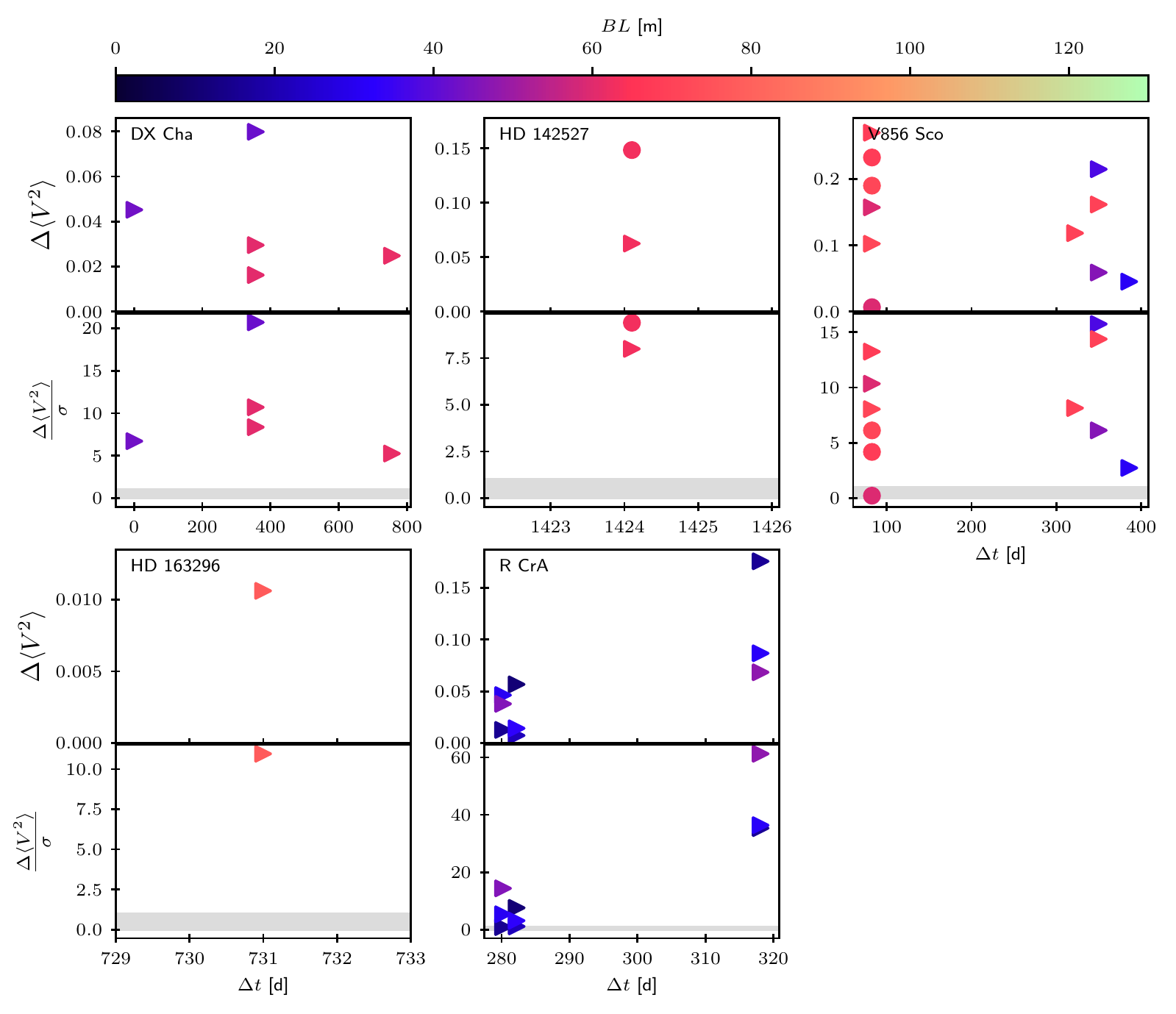}
\caption{\label{fig:sigmaAMBER} Similar to Fig.\ \ref{fig:sigmaPIONIER} but for multi-epoch observations obtained with AMBER for DX Cha, HD 142527, V856 Sco, HD 163296, and R CrA. The circles indicate differences in the squared visibilities (upper panel) as well as differences divided by the uncertainty $\sigma$ (lower panel) obtained with AMBER in the $H$ band, the triangles indicate differences in squared visibilities obtained in the $K$ band.}
\end{figure*}

In Fig.\ \ref{fig:sigmaPIONIER}, we plot the differences in the wavelength-averaged squared visibilities (upper panel of each plot) as well as the squared visibility differences expressed in multiples of the corresponding uncertainty ($\Delta \mean{V^2} / \sigma$, lower panel of each plot) against the time lag for the seven objects for which PIONIER observations obtained for equal baselines are available. In Fig.\ \ref{fig:sigmaAMBER}, we show similar plots for the AMBER observations.
From these, we can derive the following results.
We find that six of the objects (HD 50138, DX Cha, HD 142527, V856 Sco, \mbox{HD 163296}, and \mbox{R CrA}) show changes in the squared visibilities larger than $3 \sigma$, suggesting temporal variability. Variations smaller than $3 \sigma$ are not considered as significant. For example variations up to $2.8 \sigma$ in the case of CPD-36 6759 are not considered a clear indication for temporal variability.  For HD 37806 and TW Hya, only one pair of squared visibilities obtained for equal baselines is available. These show no significant differences.\\
\textbf{HD 50138:} Both pairs of squared visibilities obtained with PIONIER show a difference of $\sim\!0.01$. The more significant difference is equivalent to $5.9\sigma$. This significant deviation already occurs on the shortest timescale considered, which is two days.\\
\textbf{DX Cha:} The largest differences in the squared visibilities obtained with PIONIER is $0.25 \hat{=} 42\sigma$. The visibilities obtained in the $K$ band with AMBER differ up to $0.08 \hat{=} 21\sigma$. Therefore, significant variations already occur for the shortest covered time spans of 51 days (up to $0.23 \hat{=} 31\sigma$, PIONIER) and 1 day ($0.05 \hat{=} 6.7\sigma$, AMBER). \\
\textbf{HD 142527:} The largest difference measured with PIONIER amounts to $0.34 \hat{=} 9.5\sigma$, the difference of one pair of squared visibilities measured with AMBER both in the $H$ and $K$ band is $0.15 \hat{=} 9\sigma$ and $0.06 \hat{=} 8\sigma$, respectively. The considered PIONIER observations cover timescales as short as 1 day, for which the squared visibilities differ by $0.04 \hat{=} 1.6\sigma$.  The AMBER data only cover a time lag of 1424 days. \\
\textbf{V856 Sco:} The squared visibilities obtained with PIONIER for equal baselines with a time lag of 819 days differ by up to $0.05 \hat{=} 11\sigma$. The AMBER $H$ and $K$ band data show variations up to $0.5 \hat{=} 9\sigma$ and $0.27 \hat{=} 13\sigma$, respectively, where the latter is already found for the shortest time span covered by the AMBER data (83 days). \\
\textbf{HD 163296:} Only one pair of $K$ band visibilities has been measured with equal baselines. During the time span of 731 days between both observations, the squared visibilities changed by $0.01 \hat{=} 11\sigma$.\\
\textbf{R CrA:} $K$ band squared visibilities using equal baselines have been obtained with AMBER with time lags covering a small range from 280 to 318 days. From these, we find variations up to $0.18 \hat{=} 35\sigma$.

\subsection{Visual approach \label{sec:Plot}}
For 59 objects in our sample, multi-epoch observations have only been obtained for different baselines. In the following sections, we use different approaches to investigate these data sets for temporal variability.

\subsubsection{Description of the approach}
A first intuitive approach to analyze interferometric data is to plot all visibilities. These depend on the baseline length and orientation as well as  on the date of the observations in case of temporal variations. We plot the squared visibilities against baseline, therefore we represent each squared visibility with a bar whose orientation and color correspond to the position angle and date of the observation, respectively. For the sake of clarity and to enable a comparison of observations with different spectral resolutions, spectrally resolved observations are averaged (Eqs.\ \ref{eq:ave} - \ref{eq:sigma}). 

\begin{figure*}
\includegraphics[]{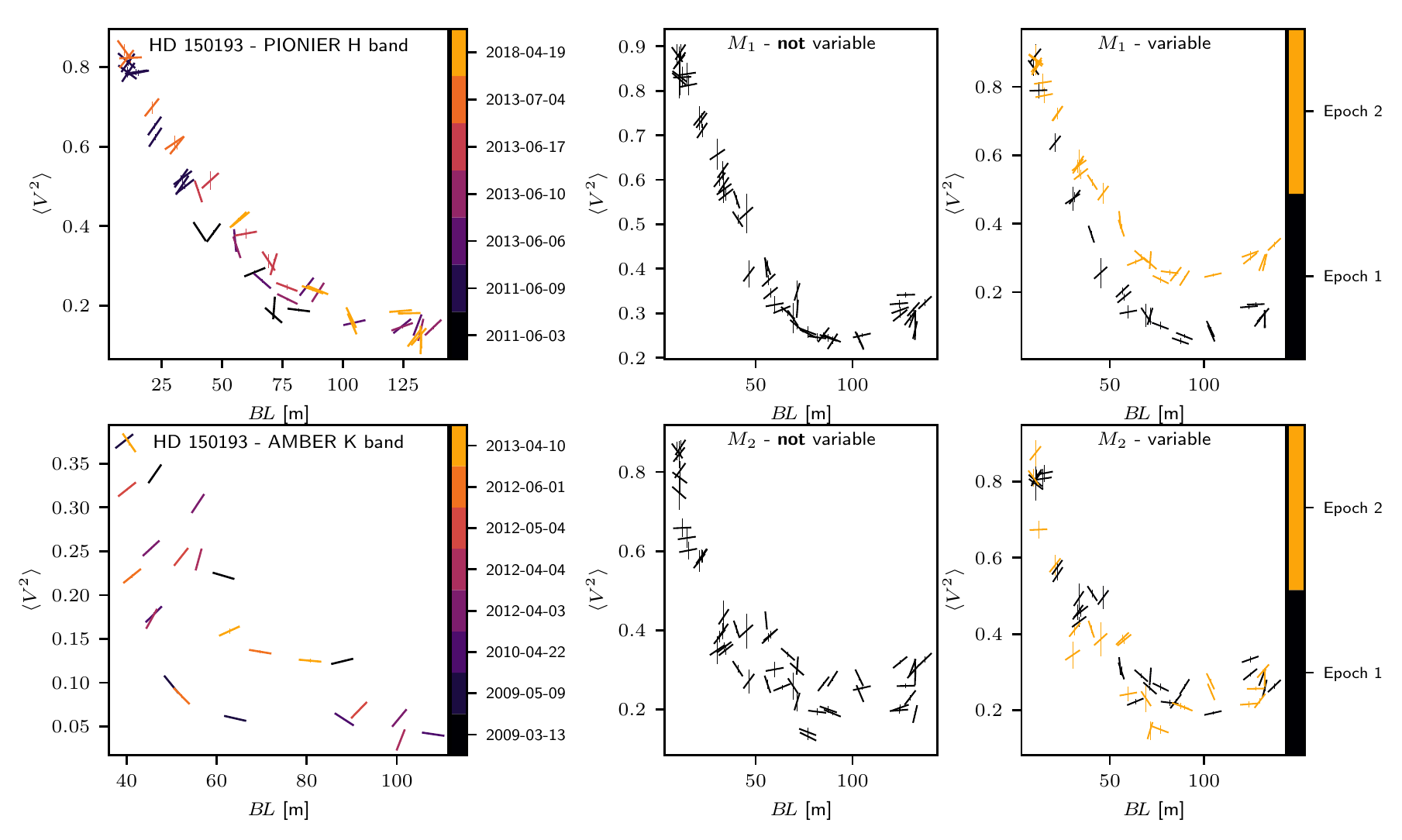}
\caption{\label{fig:illVisAp} Illustration of the visual approach (see Sect. \ref{sec:Plot}). Left column: Wavelength-averaged squared visibilities obtained with PIONIER (upper panel) and AMBER (lower panel) for HD 150193 against baseline length. Each data point is plotted as a bar indicating the position angle of the observation. The date of the observation is indicated by the color. Middle and right columns: Squared visibilities for synthetic observations based on the models $M_{1}$ (upper row) and $M_2$ (bottom row). In the middle column, the underlying brightness distribution of the synthetic data is time independent. In the right column, we varied the brightness distribution by decreasing the star-to-disk flux ratio ($M_1(t)$) or by changing the position of the disk feature ($M_2(t)$, see Sect. \ref{sec:synthObs} for details). The UV coverage as well as the error bars and the corresponding noise are adapted from the PIONIER observations of HD 150193.}
\end{figure*}

We demonstrate the visual analysis for one exemplary object, namely HD 150193 (see Fig.\ \ref{fig:illVisAp}). Here, multi-epoch observations with all three instruments are available and different aspects of the visual analysis can be demonstrated. For HD 150193, the nine considered PIONIER observations were performed on seven nights between June 2011 and April 2018 (see Table \ref{tab:logPIO} and top panel of the left column of Fig.\ \ref{fig:illVisAp}). Eight interferometric observations with AMBER were performed on eight nights between March 2009 and April 2013 (see Table \ref{tab:logAMB} and bottom panel of the left column of Fig.\ \ref{fig:illVisAp}). 
\paragraph{HD 150193 / PIONIER (Fig.\ \ref{fig:illVisAp}, left column top panel):} For every squared visibility obtained in June 2011, data obtained at later dates (June and July 2013, April 2018) using only slightly different baselines with deviations of a few meters in length and degrees in orientation exist. By comparing the measured visibilities we find that all squared visibilities from observations in 2011 are smaller than the values obtained for the nearest baseline at one of the later dates. In principle, large differences in the squared visibilities obtained at the same date can occur for small deviations of the baselines, if the observed brightness distribution is complex, for example, if there are spatially resolved structures resulting from planet-disk interaction. However, the squared visibilities obtained at baselines of similar length within one epoch show no significant deviations and the variations between different epochs show a clear trend, namely all visibilities increased from 2011 to the later dates. Thus, the variations are potentially caused by a radially symmetric change in the brightness distribution, either a symmetrical variation of the disk brightness distribution or a variation in the unresolved inner disk region (e.g., the stellar flux or the inner rim of the disk).
\paragraph{HD 150193 / AMBER (Fig.\ \ref{fig:illVisAp}, left column, bottom panel):} Comparing AMBER data from different epochs, the $K$ band squared visibilities also show significant visibility differences. The squared visibilities measured for the two short baselines ($\sim$46\,m and $\sim$62\,m) in March 2009 are significantly larger than the visibilities obtained for similar baselines in May 2009 as well as in April 2010 and  April 2012. Comparing the squared visibilities obtained with similar baseline lengths of \SI{\sim40}{\meter}, the squared visibility decreases from March 2009 to May 2012 and further to June 2012.
However, a clear trend as found in the PIONIER data does not exist. On the contrary, on 4 April 2012, the squared visibility measured for a baseline length of \SI{\sim45}{\meter} was significantly smaller than the visibility measured on the same night for the larger baseline of \SI{\sim57}{\meter} with a similar position angle, indicating a complex disk structure causing features in the brightness distribution that are spatially resolved. The differences in the $K$ band squared visibility can therefore not be clearly attributed to temporal variations in the brightness distribution.

We conclude that the PIONIER data show hints of temporal variability. The AMBER data show deviations in the squared visibilities obtained for similar baselines, which, however, cannot be clearly attributed to a temporal variation of the brightness distribution.

\subsubsection{Evaluation with synthetic observations}
We discuss the reliability of this qualitative analysis approach based on synthetic data plotted in Fig.\ \ref{fig:illVisAp}. In the upper panel of the middle column  of Fig.\ \ref{fig:illVisAp}, synthetic observations for a time-independent brightness distribution based on the model $M_{1,\mathrm{stat}}$ (simple static disk model without any features, see Sect.\ \ref{sec:synthObs}) using the UV coverage as well as the measurement uncertainties from the PIONIER observations of HD 150193 are shown. We find no large variations comparing squared visibilities obtained for baselines with similar lengths, which is expected for a nearly radially symmetric brightness distribution with a modest inclination of $30^\circ$ and no temporal variations. In the upper panel of the right column of Fig.\ \ref{fig:illVisAp}, synthetic observations based on the time-dependent model $M_1(t)$ (temporal variation of the stellar intensity, see Sect.\ \ref{sec:synthObs}) are shown. Here, the stellar intensity and thus the star-to-disk flux ratio is decreased by a factor of \num{0.5} from Epoch 1 to Epoch 2. The decrease of the star-to-disk flux ratio results in lower squared visibilities for all baseline lengths and orientations. This clear trend, together with the comparably small differences due to the baseline deviations one can estimate by comparing data obtained within an epoch, would justifiably lead to the conclusion that there is a temporal variation in the brightness distribution.

In the bottom panels of the middle and right columns of Fig.\ \ref{fig:illVisAp}, we plot synthetic data based on the models $M_{2,\mathrm{stat}}$ and $M_2(t)$ (static and time-dependent brightness distributions with a Gaussian feature at a disk radius of \SI{9}{mas}, see Sect.\ \ref{sec:synthObs}) again using the UV coverage and uncertainties from the real PIONIER observations of HD 150193. In the bottom panel of the middle column, we see that for the time-independent synthetic data, the squared visibilities obtained for apparently similar baselines show significant differences. The Gaussian disk feature induces strong variations in the squared visibility even for small deviations of the baselines.  In the lower right plot, we show synthetic data for a temporal variable brightness distribution based on model $M_2(t)$. Here, the Gaussian feature is orbiting with a period of 400 days and we choose the epochs to have a time lag of 100 days, corresponding to an orbital movement of the disk feature by \SI{90}{\degree}. The resulting differences in the squared visibilities are slightly larger than those that are solely caused by the brightness asymmetry (lower left plot). A clear trend cannot be seen due to the asymmetric variation of the brightness distribution. 
Thus, it is not possible to distinguish clearly between the occurrence of brightness asymmetries and temporal variations without estimating the effect of small baseline deviations on the observed visibilities. However, this requires sufficient UV coverage with similar baselines on several spatial scales and orientations. 

We conclude that it is not possible to unambiguously identify the cause of differences on the basis of a qualitative evaluation with the plots presented. Nevertheless, these plots can be helpful to identify candidates for temporal variability. However, a more sophisticated analysis of the data, including quantitative estimates of the impact of baseline variations on the squared visibilities, is necessary to constrain temporal variations of the brightness distribution.

\subsection{Enhanced comparison approach \label{sec:extComparison}}
In Sect. \ref{sec:dircomp}, we compared data obtained at different epochs using equal baselines, whereby significant deviations in the squared visibilities were considered as clear indications of temporal variations in the brightness distribution. In the following, we extend the comparison of squared visibilities measured at different epochs to pairs of observations with moderately deviating baselines. In this case, the resulting differences can be caused by both, a) temporal variations in the brightness distribution and b) baseline deviations being sensitive to the complex structure of a brightness distribution. To distinguish these two causes, we estimate the impact of the baseline deviation by comparing observations obtained within one epoch.

\subsubsection{Description of the approach}
We proceed as follows. We first calculate the following quantities for each combination of two measured visibilities: 
a) The absolute differences of the wavelength-averaged squared visibilities $|\mean{V_k^2} - \mean{V_j^2}|$, 
together with the corresponding uncertainty $\sigma$ as described in Eqs. (\ref{eq:ave}) to (\ref{eq:sigma}); b) the relative baseline deviation
\begin{equation}
b = 2\frac{\sqrt{\left(u_k - u_j\right)^2 + \left(v_k - v_j\right)^2}}{\mathit{BL}_j + \mathit{BL}_k},
\end{equation}
given by the ratio of the Euclidean distance of the two baselines to their mean length. We consider all observations with moderate baseline differences of $b<10\,\%$; c) the time lag $\Delta t = t_k - t_j$ between epochs $j$ and $k$. 

\begin{figure*}
\includegraphics[width=\textwidth]{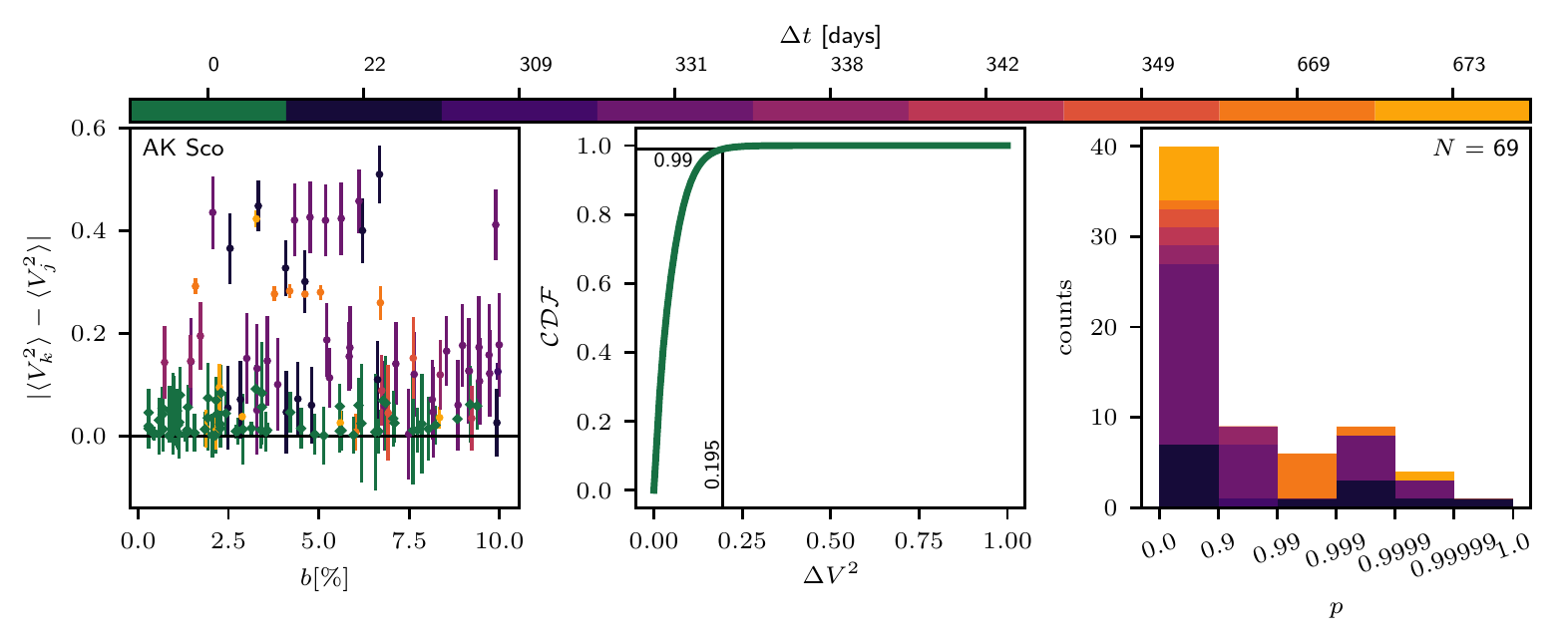}
\caption{Illustration of the analysis approach based on the comparison of visibilities taking into account estimates for typical variations due to moderate baseline deviations as described in Sect.\ \ref{sec:dircomp} by the example of AK Sco. Left: Differences of all pairs of squared visibilities obtained using baselines with a maximum difference of \SI{10}{\percent} with PIONIER. Differences calculated for two squared visibilities obtained within the same night are plotted as green diamonds. The points represent differences calculated for two squared visibilities measured on different nights with a time lag $\Delta t$ indicated by the color bar.  Middle: Cumulative distribution function calculated from the differences of the squared visibilities with $\Delta t < 12\,$h (Eq. \ref{eq:cdf}). The black lines indicate the probability of 0.99 to obtain a visibility difference below $0.195$ when expecting the difference to be affected solely by the baseline deviation and not by temporal variability. Right: Histogram of the measure $p$ (Eq. \ref{eq:p}), indicating the probability of the 69 visibility differences with $\Delta t > 12\,h$ ($\Delta t$ indicated by the color) to be affected by temporal variations in the brightness distribution.
\label{fig:DeltaVis2VSDeltaBL}}
\end{figure*}

In order to illustrate the idea of this approach, as an example we plot the differences for the comparison of all visibilities obtained with PIONIER for AK Sco against the relative baseline deviation $b$ and thereby color-code the time lag between the observations (see Fig \ref{fig:DeltaVis2VSDeltaBL}). Assuming that the brightness distribution does not change significantly within one night, we can estimate the typical impact on the visibilities of changing the baseline by up to ten percent by inspecting the visibility differences with $\Delta t < 12$\,h, plotted with green diamonds in the left panel of Fig. \ref{fig:DeltaVis2VSDeltaBL}.
However, temporal variations of the brightness distribution can occur even in a few hours (see Sect. \ref{sec:introduction}). In this case we would erroneously attribute the corresponding visibility differences to the baseline deviation. Consequently, this approach is not sensitive to temporal variability on timescales of less than 12 hours. However, for a reasonable estimate of the impact of baseline variations, a sufficient number of differences calculated from visibilities obtained within one epoch are required, which cannot be achieved when only comparing observations with time lags significantly shorter than 12 hours.
For AK Sco, we find that the differences in the squared visibilities induced by small baseline deviations of less than ten percent are typically below $0.2$. Comparing the differences with time lags between 22 and 673 days, we find significantly larger values of up to $|\mean{V_k^2} - \mean{V_j^2}| = 0.54$, suggesting that the brightness distribution of AK Sco is temporally variable.

To quantify this analysis, we calculate a measure for the probability that the difference of two squared visibilities is larger than we would expect to result from the given baseline deviation alone. For this purpose, we assume that every difference $|\mean{V_k^2} - \mean{V_j^2}|$ with an uncertainty $\sigma$ (see Eq.\ \ref{eq:sigma}) is represented by a normally distributed probability density function $\mathit{pdf}$. Since we calculate the absolute difference, the $\mathit{pdf}$ must only provide positive values. We therefore add the part of the probability density function for negative differences to the probability of the corresponding positive differences:

\begin{align}
        \mathit{pdf}_{j,k}\left(\Delta V^2\right) &\propto& \frac{1}{\sqrt{2 \pi} \sigma} \exp{\left(-\frac{\left(\Delta V^2-|\mean{V_k^2} - \mean{V_j^2}|\right)^2}{2 \sigma^2}\right)} \nonumber \\&&+  \frac{1}{\sqrt{2 \pi} \sigma} \exp{\left(-\frac{\left(\Delta V^2+|\mean{V_k^2} - \mean{V_j^2}|\right)^2}{2 \sigma^2}\right)}. \label{eq:pdf}
\end{align}
Summing the probability density functions of $N$ pairs of squared visibilities measured within the same epoch ($\Delta t < 12$\,h), we obtain the probability density function for the differences of the squared visibilities $\mathcal{PDF}$ we expect due to a variation of the baseline,
\begin{align}
\mathcal{PDF}\left(\Delta V^2\right) = \frac{\sum_{i=1}^{N} \mathit{pdf}_{\Delta t < 12\,\mathrm{h},\ i}\left(\Delta V^2\right)}{\int_0^1\sum_{i=1}^{N} \mathit{pdf}_{\Delta t < 12\,\mathrm{h},\ i}\left(\Delta V^2\right) \mathrm{d}\Delta V^2}.
\end{align}
We presuppose that at least ten pairs of squared visibilities obtained with $\Delta t < 12\,\mathrm{h}$ using moderately different baselines ($b<10\,\%$) must be available in order to be able to reasonably estimate the impact of the baseline deviation on the squared visibilities. Pre-main sequence stars in our sample for which the existing observations do not meet this criterion are excluded from this analysis. From the $\mathcal{PDF}$, we can calculate the cumulative distribution function 
\begin{align}
\mathcal{CDF}\left(\Delta V^2\right) = \int_0^{\Delta V^2} \mathcal{PDF}\left(\widetilde{\Delta V^2}\right) \mathrm{d}\widetilde{\Delta V^2}. \label{eq:cdf}
\end{align}
This gives us the probability of obtaining a difference lower than a given $\Delta V^2$ that is caused by a moderate baseline difference (and not by temporal variations).

From the $\mathcal{CDF}$ calculated for AK Sco (see middle panel of Fig. \ref{fig:DeltaVis2VSDeltaBL}), we find that visibility differences caused by moderate baseline deviations ($b<\SI{10}{\percent}$) are less than \num{0.195} with a probability of \SI{99}{\percent}. Obtaining a difference of two squared visibilities larger than $0.195,$ which is not caused by temporal variations of the brightness distribution, thus has a probability of only $1\,\%$.
In order to estimate whether a difference of two squared visibilities obtained in two different epochs ($\Delta t>12\,$h) is larger than expected to result from the baseline deviation (i.e., affected by temporal variations), we calculate the measure:
\begin{align}
p = \int_0^1 \mathcal{CDF}\left(\Delta V^2\right)\ \mathit{pdf}_{\Delta t \geq 12\,\mathrm{h}; j,k}\left(\Delta V^2\right) \mathrm{d}\Delta V^2. \label{eq:p}
\end{align}
The quantity $p$ has values between zero and one, where the probability that a change of the squared visibility is affected by temporal variations of the brightness distribution is the higher the larger $p$.
For the largest visibility difference found for AK Sco of $\Delta V^2 = 0.54$ this measure amounts to $p=0.9999985$, indicating a high probability for this difference to be caused by temporal variations in the brightness distribution.
Of a total of 69 pairs of squared visibilities obtained for AK Sco in different epochs ($\Delta t > \SI{12}{\hour}$) with moderately different baselines, 16 have a difference that corresponds to a value of $p$ greater than \SI{99.9}{\percent} (see right panel of Fig.\ \ref{fig:DeltaVis2VSDeltaBL}). For three differences, the value of $p$ even exceeds $99.999\,\%$.

Eventually, we calculate a measure for the total probability that an object is temporally variable, taking into account all differences obtained in different epochs using moderately different baselines. Especially in the case of many measurements, we expect the occurrence of unlikely events to a small extent. This means that for a large number of considered visibility differences we expect to obtain a small number of those for which we individually infer a high probability of temporal variability ($p > \SI{99}{\percent}$). On the other hand, it is possible that a large fraction of differences, each with only a moderate probability ($\SI{90}{\percent} < p < \SI{99}{\percent}$) and thus individually not considered to be caused by temporal variability, indicate temporal variability.

For a total number of $N$ differences ( $|\mean{V_k^2} - \mean{V_j^2}|$ with $\Delta t>12\,$h and $b<\SI{10}{\percent}$) we calculate the probability to obtain at least $N_{p>p_\mathrm{lim}}$ differences that are solely caused by the moderate baseline deviation with a value of $p$ exceeding a given limit $p_\mathrm{lim}$:

\begin{align}
\mathcal{P}\left( X \geq N_{p>p_\mathrm{lim}} \right) \leq \sum_{k=N_{p>p_\mathrm{lim}}}^{N} \binom{N}{k}\ p_\mathrm{lim}^k \left(1-p_\mathrm{lim}\right)^{N-k}. \label{eq:P}
\end{align}
The quantity $1-\mathcal{P}\left( X \geq N_{p>p_\mathrm{lim}} \right)$ is therefore a measure of the probability that visibilities obtained in different epochs suggest temporal variations of the brightness distribution.
For 16 differences with $p>99.9\,\%$ out of a total of 69 considered differences (see right panel of Fig.\ \ref{fig:DeltaVis2VSDeltaBL}) this value amounts to $\mathcal{P}=3.04 10^{-33}$. The visibility differences found for AK Sco are therefore unlikely to be caused solely by the moderate baseline deviations. Therefore, the brightness distribution of AK Sco is temporally variable with a high probability.

For a quantitative evaluation of the evidence of temporal variability, we calculate the probability measure $\mathcal{P}$ for different limits $p_\mathrm{lim} \in \left\lbrace 0.9, 0.99, 0.999, 0.9999, 0.99999 \right\rbrace$. If $\mathcal{P}$ is below $0.001$ (indicating a high probability of temporal variability with $1-\mathcal{P}> 99.9\,\%$) for at least one of these limits, we conclude that the archival interferometric data show evidence for temporal variations in the brightness distribution.

\subsubsection{Evaluation with synthetic observations}
To evaluate the reliability of the above approach, we apply it to synthetic data simulated for the different models with and without a variable brightness distribution. First, we calculate synthetic data for the two static models $M_{1,\mathrm{stat}}$ and $M_{2,\mathrm{stat}}$ (see Sect..\ \ref{sec:synthObs}) using the UV coverage, dates, and error bars of PIONIER observations obtained for 18 objects (those with at least ten pairs of observations with $\Delta t < 12\,\mathrm{h}$ and $b<10\,\%$). Applying the approach to analyze these time-independent 36 synthetic data sets did not indicate variability in any of them. For the model $M_{1,\mathrm{stat}}$ (static brightness distribution with only a slight asymmetry due to the disk inclination of \SI{30}{\degree}, see Sect.\ \ref{sec:synthObs}), we find the most stringent values of $\mathcal{P}$ (indicating the probability that the visibility differences are solely caused by the baseline deviations, see Eq.\ \ref{eq:P}) to be between \num{0.92} and \num{1}. From this we properly conclude that the underlying brightness distribution is not variable. 

For the static model $M_{2,\mathrm{stat}}$ (Gaussian disk feature, see Sect.\ \ref{sec:synthObs}), we find the lowest value of $\mathcal{P} = 0.26$ for the synthetic data based on the UV and temporal coverage of the PIONIER observations of HD 179218. A higher probability for variability ($1-\mathcal{P} = 0.74$) as the cause of the visibility differences for model $M_{2,\mathrm{stat}}$ compared to $M_{1,\mathrm{stat}}$ can be explained by the limited UV coverage together with the asymmetry of the brightness distribution. The impact of the asymmetry on the visibility depends on the specific spatial frequency, that is, the chosen baseline. Visibility differences may thus be due to even moderate baseline variations. Dissimilar UV coverage regarding the pairs of visibilities obtained within one and different epochs may thus lead to an erroneous misidentification of the differences as caused by variability.
However, a probability as large as $\mathcal{P} = 0.26$ significantly exceeds our limit (\num{0.001}) for the identification of temporal variability. Moreover, it is very unlikely that we would obtain values of $\mathcal{P} < 0.001$ due to the described effect. 

Subsequently, we simulate synthetic observations with variable brightness distributions based on the models $M_1(t)$ and $M_2(t)$ again using the UV coverage and temporal coverage of the 18 PIONIER data sets. For the synthetic data based on the model $M_1(t)$ (variation of the star-to-disk flux ration, see Sect.\ \ref{sec:synthObs}), we are able to identify variations for 10 of the 18 data sets with this approach. In the case of the synthetic observations based on model $M_2(t)$ (Gaussian disk feature with an orbital period of 400 days, see Sect.\ \ref{sec:synthObs}), we find $\mathcal{P}< 0.001$ suggesting temporal variability for 5 of the 18 synthetic data sets. We discuss the different reasons for a non-detection of variability in Sect.\ \ref{sec:discussion}.

We conclude that for multi-epoch observations with sufficient UV and temporal coverage this approach is suitable for detecting variability. Moreover, our performed tests indicate a low probability of obtaining false-positive results.

\subsubsection{Results}

The presented approach based on the comparison of visibilities taking estimates for typical variations due to moderate baseline deviations into account can be used to investigate temporal variability based on PIONIER data for 18 of the pre-main sequence stars in our sample (GW Ori, CQ Tau, HD 50138, TW Hya, DX Cha, CPD-36 6759, HD 142527, RU Lup, HD 144432, V856 Sco, AS 205 N, SR 21A, HD 150193, AK Sco, HD 163296, HD 169142, S CrA N, and HD 179218). Based on AMBER data, this analysis is possible for seven objects (HD 50138, CPD-36 6759, HD 142527, V856 Sco, HD 163296, MWC 297, and R CrA),  five of which are also part of the list of PIONIER targets. Due to the sparse UV coverage, this analysis cannot be performed for any of the MIDI observations of the considered objects. The number of visibility differences that can be calculated for observations with $\Delta t < \SI{12}{\hour}$ is not sufficient to provide a useful estimate of the impact of the baseline variation. Therefore, the approach is suitable to analyze the archival interferometric data obtained for 20 objects in our sample.

We find evidence for temporal variations for 6 out of these 20 pre-main sequence stars. In Table \ref{table:A4}, we list the instrument and spectral band used to obtain these data, with the most stringent value of the quantity $\mathcal{P}$ (indicating the probability that the visibility differences are caused solely by the baseline deviations, Eq.\ \ref{eq:P}) calculated for the different limits $p_\mathrm{lim} \in \left\lbrace 0.9, 0.99, 0.999, 0.9999, 0.99999 \right\rbrace$, the largest difference of the squared visibilities with the corresponding value of $p$ (Eq.\ \ref{eq:p}), as well as the shortest timescale that has visibility differences with $p>0.9$.

For HD 50138, DX Cha, V856 Sco, HD 163296, and R CrA the direct comparison of visibilities obtained for similar baselines has already revealed evidence for variability (see Figs.\ \ref{fig:sigmaPIONIER} and \ref{fig:sigmaAMBER} and the discussion in Sect.\ \ref{sec:dircomp}). For HD 50138 and HD 163296, variability was previously identified using PIONIER and AMBER data, respectively. Now we find signs of temporal variability based on data obtained for the other respective instrument as well. Expanding the comparison to squared visibilities that have been obtained for baselines differing by up to \SI{10}{\percent}, we confirm the previous findings with $\mathcal{P} = \num{9.8e-4}$, $\mathcal{P} = \num{1.36e-329}$, $\mathcal{P} = \num{3.08e-11}$, $\mathcal{P} = \num{3.84e-49}$, and $\mathcal{P} = \num{2.01e-13}$, respectively. 

As we take more data into account, we have a better temporal coverage for this approach. With that, we now already find evidence for temporal variations on shorter timescales for V856 Sco (1 day), HD 163296 (6 days), and R CrA (1 day). However, these are again the shortest timescales covered, thus the temporal variations found for these pre-main sequence stars may occur on even shorter time spans. Based on the direct comparison, for DX Cha we found variability on timescales of 51 days and one day with PIONIER and AMBER, respectively. Here we find variations on the shortest considered timescale of one day for both instruments. Only for HD 50138 do the AMBER data considered here reveal a longer timescale for temporal variations (56 days) than the direct comparison of the PIONIER data (two days).

We now find evidence for temporal variations of the brightness distributions of one additional object. For AK Sco the largest difference in the squared visibilities in $H$ band is 0.54 (compare Fig. \ref{fig:DeltaVis2VSDeltaBL}). The corresponding $p = 0.9999985$ indicates a high probability that this difference is caused by temporal variations of the brightness distribution. With 16 differences having a value of $p > \SI{99.9}{\percent}$ out of 69 differences obtained from multi-epoch observations with a baseline deviation below \SI{10}{\percent}, the PIONIER observations show strong indications of temporal variations with $\mathcal{P} = \num{3.04e-33}$. We already find differences that are probably affected by temporal variations for the shortest covered time span of 22 days.

\setcounter{table}{1}
\begin{table*}
\renewcommand{\arraystretch}{1.2}
\caption{Overview of the objects showing strong indications of temporal variability based on the comparison approach including multi-epoch observations with baseline deviations up to \SI{10}{\percent} (see Sect. \ref{sec:extComparison} for details). Objects, for which the shortest timescale with significant variations corresponds to the shortest timescale covered by the observations, are marked with a star $^\star$.}             
\label{table:A4}      
\centering          
\begin{tabular}{l c c c c c c}    
\hline\hline       
Name & Instrument & Band & $\mathcal{P}_\mathrm{min}$ & $\Delta V^2_\mathrm{max}$ & $p\left(\Delta V^2_\mathrm{max}\right)$ & $\Delta t_\mathrm{var}$ [d]\\ 
\hline          
HD 50138        & AMBER         & K & \num{9.8e-4}              & \num{0.318}   & \num{0.999999999996}    & $56$ \\
DX Cha          & PIONIER       & H & \num{1.36e-329}   & \num{0.281}   & \num{0.999999999985}    & $1^\star$\\    \vspace{-4pt}
\multirow{2}{*}{V856 Sco}       & PIONIER       & H & \num{3.08e-11}    & \num{0.156}     & \num{0.997}                   & $1^\star$\\           
                        & AMBER         & K & \num{1.65e-7}             & \num{0.531}     & \num{0.999999991}             & $1^\star$\\ 
AK Sco          & PIONIER       & H & \num{3.04e-33}    & \num{0.539}   & \num{0.9999985}         & $22^\star$\\ \vspace{-4pt}
\multirow{2}{*}{HD 163296}      & PIONIER       & H & \num{2.95e-36}    & \num{0.124}     & \num{0.99895}                 & $12$ \\               
                        & AMBER         & K & \num{3.84e-49}    & \num{0.51}    & \num{1.0}                       & $6^\star$ \\
R CrA           & AMBER         & K & \num{2.01e-13}    & \num{0.464}   & \num{0.99999998}                & $1^\star$ \\
\hline                  
\end{tabular}
\end{table*}

%================================================================================
%================================================================================
\section{Discussion\label{sec:disc}}
%================================================================================
We have presented two approaches (direct comparison approach and a comparison approach including estimates of moderate baselines variations, see Sects.\ \ref{sec:dircomp} and \ref{sec:extComparison}) that are suitable for studying the temporal variability of pre-main sequence stars. Based on these approaches, we find that 7 of the 68 objects in the sample show evidence for temporal variability. For these objects, we discuss previous variability studies in Sect.\ \ref{sec:previous_studies}. In Sect.\ \ref{sec:discussion} we discuss the reasons for the non-identification of temporal variations. Finally, in Sect.\ \ref{sec:trend} we classify the detected variations in order to allow conclusions about their potential origin.

%================================================================================
%================================================================================
\subsection{Known temporal variability \label{sec:previous_studies}}
%================================================================================
 We find indications of temporal variations for the seven pre-main sequence stars HD 50138, DX Cha, HD 142527, V856 Sco, AK Sco, HD 163296, and R CrA. However, for all of these objects, variability was either already found in previous studies or was expected, as a companion is present whose orbital motion causes variations in the brightness distribution.\\
\textbf{HD 50138:} Spatially unresolved observations already revealed the temporal variability of several spectral lines in the UV and visible on timescales from hours to years \citep{2012A&A...548A..13B, 1997A&A...317..185P, 1985A&AS...60..373H}. In a recent study \citet{2016A&A...591A..82K} investigated the near-infrared variability based on images reconstructed from PIONIER observations obtained in three different epochs and found strong morphological changes in the innermost disk region on a timescale of a few months.\\
\textbf{DX Cha:} As DX Cha is a spectroscopic binary with a period of 20\,days \citep{2004A&A...427..907B}, whose semi-major axis of $0.2\,$au (\SI{1.9}{mas}) is at least partially resolved by the PIONIER and AMBER observations, we expect variations due to the orbital motion of the binary. In addition, \citet{2012ApJS..201...11K} found wavelength-independent variations when comparing four mid-infrared spectra obtained between July 1996 and May 2005.\\
\textbf{HD 142527:} The transitional disk of HD 142527 contains a low-mass stellar companion at \SI{\sim12}{au} from the star \citep{2014ApJ...781L..30C, 2012ApJ...753L..38B}. Due to the large star-to-companion flux ratios \citep[$H$ band: 63, $Ks$ band: 79; ][]{2016A&A...590A..90L} we expect only small variations due to the orbital movement of the companion. Assuming a model consisting of two point sources, we can estimate the impact of the companion on the squared visibilities to $\lesssim 0.06$.\\
\textbf{V856 Sco:} Analyzing a light curve of V856 Sco spanning 19 years, \citet{1992A&A...257..209P} found variations of the visual magnitude between $6.74$ and \SI{8.46}{mag}, probably caused by variable circumstellar extinction.\\
\textbf{AK Sco:} Ak Sco is a spectroscopic binary with a period of 13.6 days \citep{1989A&A...219..142A} and a projected separation of \SI{0.143}{au} \citep{1996ApJ...458..312J}.  On the one hand, we expect variations due to the orbital motion of almost identical stars \citep{2003A&A...409.1037A}. In addition, variations in optical \citep{1991A&AS...89..319B} and infrared \citep{1994A&A...285..883H} wavelengths have been reported that are not related to the orbital period. To explain the opposite behavior of the variations in the different wavelength ranges, \citet{1994A&A...285..883H} proposed variations of the density distribution at radii larger than \SI{0.6}{au}.\\
\textbf{HD 163296:} For the Herbig Ae star HD 163296, numerous studies have already revealed temporal variability. Variations with a period of 16 years due to mass ejection have been found in optical and infrared observations \citep{2014A&A...563A..87E, 2008ApJ...678.1070S}. Here, the contrary behavior, a fading in the optical and at the same time brightening infrared flux, are indicative of dust clouds in the wind \citep{2014A&A...563A..87E}. Furthermore, images revealed a time-dependent scattered light morphology in the outer disk region of HD 163296 \citep{2019ApJ...875...38R, 2008ApJ...682..548W}. Dust clouds ejected above the disk midplane, partially shading the outer disk regions, are also a possible explanation for this phenomenon \citep{2019ApJ...875...38R}.\\
\textbf{R CrA:} The light curve of R CrA shows a long-term variation with a decrease starting at ~10 mag to a magnitude oscillating between 12 and 13,  potentially caused by a progressive increase in the disk absorption \citep[based on data from the American Association of Variable Star Observers, ][]{2019A&A...630A.132S} . In addition, the light curve of R CrA shows variations with a period of 66 days \citep{2010JAVSO..38..151P}. \citet{2019A&A...630A.132S} were able to model these variations using a binary with a semi-major axis of \SI{0.56}{au} hosting a circumbinary disk (inclination \SI{>70}{\degree}). Here, the variations are due to the fact that parts of the orbit are hidden by the near side of the disk and thus the stellar light on parts of the orbit is obscured by the circumbinary material.

We conclude that for all objects for which we identify temporal variations, this behavior was expected. However, at the same time we do not identify variability for some of the objects for which it was expected. In particular, these are HD 50138 and DR Tau, introduced in Sect.\ \ref{sec:introduction}, for which interferometric observations - which are also considered in this study - have revealed variability. In the case of DR Tau, this is because the approaches presented are not suitable. There are no multi-epoch observations with equal baselines that allow the direct comparison presented in Sect.\ \ref{sec:dircomp}. Furthermore, there are not enough observations obtained within one epoch using similar baselines that would allow us to perform the analysis described in Sect.\ \ref{sec:extComparison}  based on the estimation of the impact of small baseline deviations. For HD 50138, both the direct comparison of PIONIER observations obtained with equal baselines (see Sect.\ \ref{sec:dircomp}) and the comparison of AMBER observations, taking into account expected variations due to small baseline deviations (see Sect.\ \ref{sec:extComparison}), revealed temporal variability. However, we had expected that the latter approach would also have shown variations for the PIONIER data. Possible reasons for this non-detection are discussed subsequently (Sect.\ \ref{sec:discussion}). Similarly, the brightness distribution of some binaries was not recognized as variable. For T Tau S, GG Tau, and Z CMa, for which only MIDI data is available, the analysis based on the presented approaches is not possible at all. However, for GG Tau the comparison of PIONIER observations, taking into account expected variations due to small baseline deviations, is possible, but we could not identify any temporal variations.

We conclude that the available data together with the presented approaches do not allow us to identify existing variability in several cases. However, the fact that variability was expected for all objects identified as variable and that we did not obtain false-positive results in the analysis of synthetic data, make us confident that the approaches presented provide reliable indications of temporal variability.

%================================================================================
%================================================================================
\subsection{Implications of non-detected variability \label{sec:discussion}}
%================================================================================
For many objects we could not detect temporal variability, although some of them have been found to be variable already in earlier studies or are expected to show variations due to the orbital motion of companions. In the following, we discuss the prerequisites for detecting variability based on the available data with the presented methods.

\begin{enumerate}
        \item[a)] The configuration of the interferometric observations must be such that at least one of the two approaches to investigate temporal variability (see Sect.\ \ref{sec:dircomp} and Sect.\ \ref{sec:extComparison}) is applicable. This is the case for PIONIER and/or AMBER data obtained for only 21 of the 68 pre-main sequence stars in our sample.
        \item[b)] The sensitivity of the interferometric measurements has to be better than the visibility variations induced by temporal changes in the brightness distribution. Variations of the brightness distributions thus can only be detected if they induce changes in the visibilities larger than the respective measurement uncertainties. In particular, the measured visibilities depend on the brightness distribution, but not on the total amount of flux. As a consequence, even strong variations in the total flux, which can easily be observed with unresolved photometric observations, cannot be detected by the interferometric observations if the brightness distribution does not change significantly.
        \item[c)] The temporal coverage of the observations has to cover the timescales on which the brightness distribution is variable. In the approach presented in Sect.\ \ref{sec:extComparison}, we use all observations with a $\Delta t < \SI{12}{\hour}$ to estimate the impact of small variations of the baseline. Thus, we are not sensitive to variations on timescales of a few hours. Furthermore, this procedure even has the consequence that variations on longer timescales cannot be detected if the differences due to the long-term variations are not significantly larger than those due to the variations occurring on short timescales, which are erroneously attributed to baseline variations as the cause.
        \item[d)] The UV coverage has to include the spatial scales on which the variation of the brightness distribution is significant. For example the orbital motion of a structure in the disk can only be traced if the baselines are long enough to at least partly spatially resolve the asymmetry.
\end{enumerate}
In summary, 21 pre-main sequence stars in our sample fulfill the criteria allowing us to constrain temporal variability in the near-infrared based on the approaches presented in Sects.\ \ref{sec:dircomp} and\ \ref{sec:extComparison}. Of these, seven show evidence of temporal variations. The fact that we do not find evidence for variable brightness distributions in the other 14 objects does not necessarily mean that they are actually not variable. The number of variable objects is therefore expected to be significantly larger than our results suggest.

In order to optimize interferometric observations for variability studies, future observations should be scheduled in such a way that equal baselines to those already covered in earlier epochs are used. In this case, one can directly compare the visibilities measured in different epochs, whereby any differences can be directly attributed to temporal variations in the brightness distribution (as presented in Sect.\ \ref{sec:dircomp}). In addition, snapshots with a sufficient UV coverage allowing us to perform image reconstructions could provide the basis for studying the disk structure and potential asymmetries causing variable brightness distributions.

%================================================================================
%================================================================================
\subsection{Classification of the variations \label{sec:trend}}
%================================================================================
We aim to constrain the origin of the variations found for the seven objects. One approach that is often used to investigate the causes of variations found in spatially unresolved observations is an analysis of the periodicity and the timescales on which the variations occur. However, this is not possible with the present interferometric observations due to the sparse temporal coverage. Instead, we take advantage of the spatial resolution of the observations to classify the variations.

In Sect.\ \ref{sec:Plot} we found that we can distinguish two different types of variability based on their observational appearance. 1.) Symmetric variations. Here, a clear trend can be seen, namely that the visibilities all either increase or decrease independent of the position angle ($M_\mathrm{1}$, upper right panel of Fig.\ \ref{fig:illVisAp}). With the term 'symmetric' we refer to the sign of the variation only, meaning that depending on the position angle the magnitudes of the variations can be quite different. 2.) Asymmetric variations. In this case we find both increasing and decreasing visibilities for different position angles ($M_\mathrm{2}$, lower right panel of Fig.\ \ref{fig:illVisAp}).

First, we have to relate these two variability types, different in their appearance, to the physical origins described in the introduction (cool and hot spots, obscuration of the stellar flux by circumstellar material, orbital motion of asymmetries in the brightness distribution). In the case of an orbital motion of an asymmetry in the brightness distribution, position angles from which the asymmetry moves away provide contrary variations in the visibilities, as those toward which the asymmetry moves. The visibilities measured for different position angles thus both increase and decrease, resulting in the variations being classified as asymmetric. The obscuration of the stellar flux by circumstellar material results primarily in a variation of the star-to-disk flux ratio causing a symmetrical variation. However, it may also occur that the asymmetry in the density distribution causing the darkening results in a significant asymmetry in the disk brightness distribution. In this case its movement would cause an asymmetric variation. Stellar spots cause variations both in the stellar flux and the disk brightness distribution due to the changed heating and the variations of the scattered stellar light. Here, the asymmetries in the disk brightness distribution caused by the irregular distribution of the spots may be negligible. In this case the stellar spots cause symmetrical variations. Alternatively, the resulting asymmetries in the brightness distribution are significant. In this case we classify the variations as asymmetric.

In summary, symmetrical variations are caused by changes in the stellar flux, which result in variations of the star-to-disk flux ratio and also in variations of the disk brightness caused by the changed heating and the variations of the scattered stellar light.
Asymmetric variations are caused either by companions, asymmetries in the dust density distribution, or asymmetric illuminations of the circumstellar material, meaning they result from stellar spots or obscuration.

\begin{figure*}
\centering
\newlength{\figwidthtrend}
\setlength{\figwidthtrend}{0.48\textwidth}
 \includegraphics[width=\figwidthtrend]{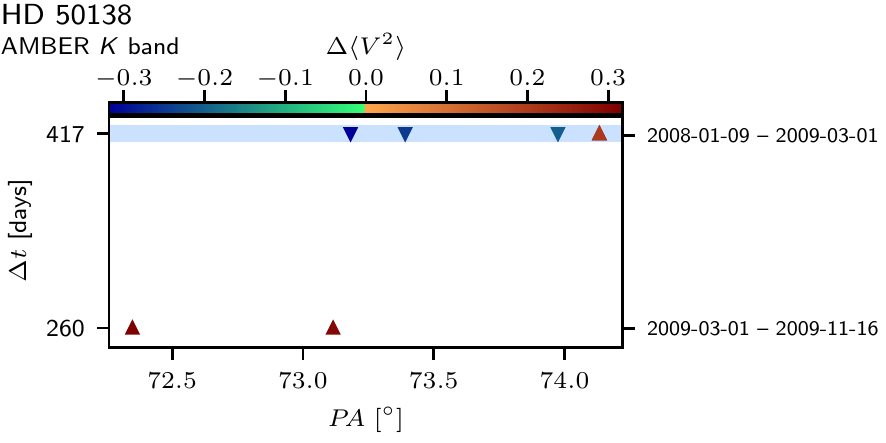}
 \includegraphics[width=\figwidthtrend]{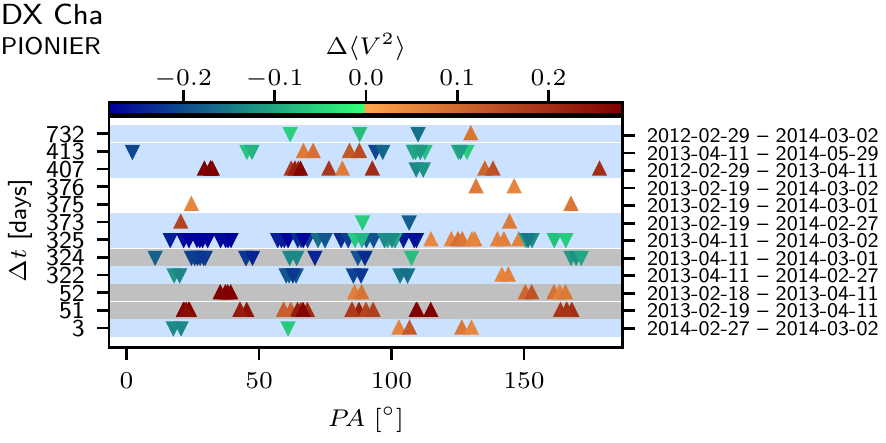}

 \includegraphics[width=\figwidthtrend]{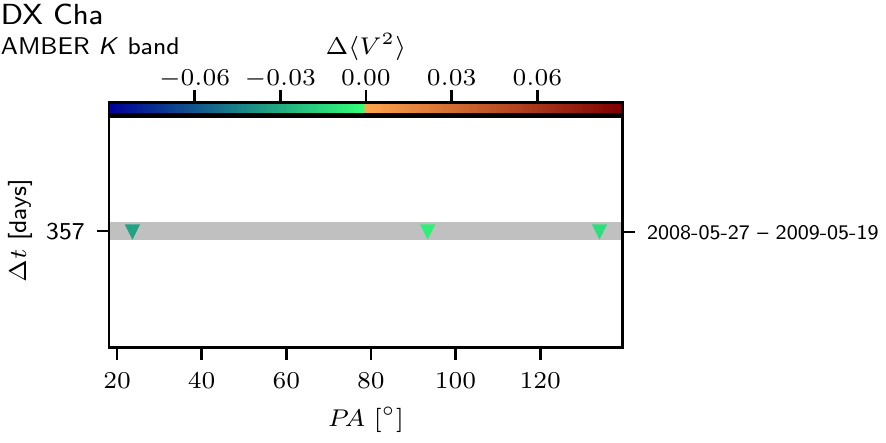}
 \includegraphics[width=\figwidthtrend]{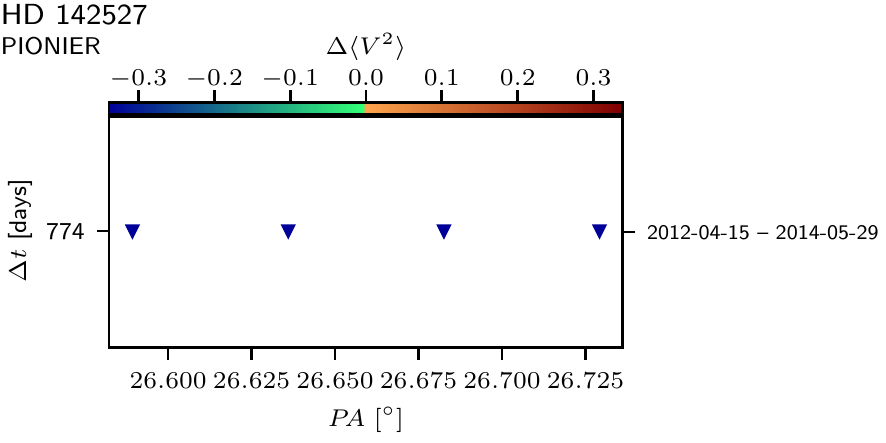}

 \includegraphics[width=\figwidthtrend]{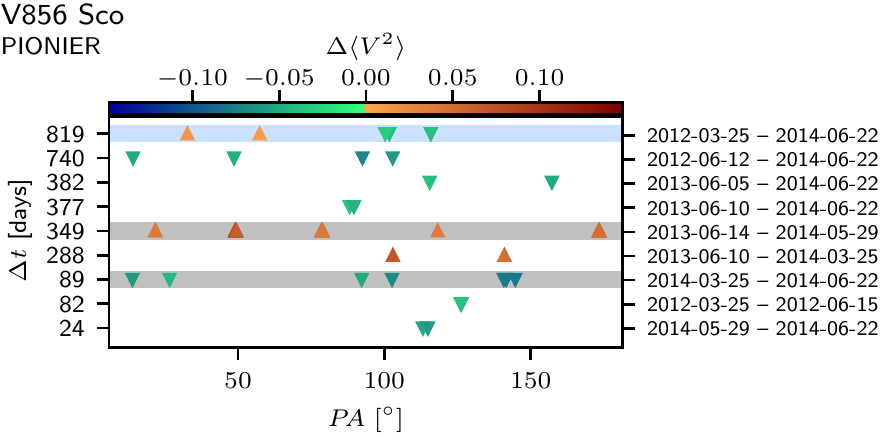}
 \includegraphics[width=\figwidthtrend]{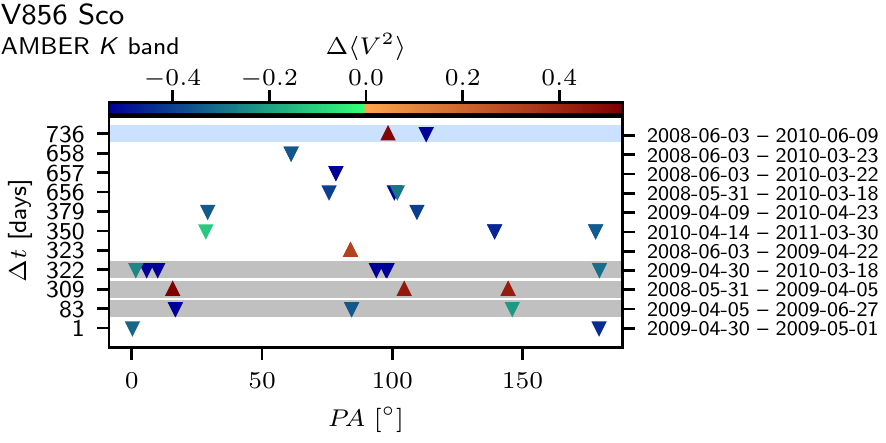}

 \includegraphics[width=\figwidthtrend]{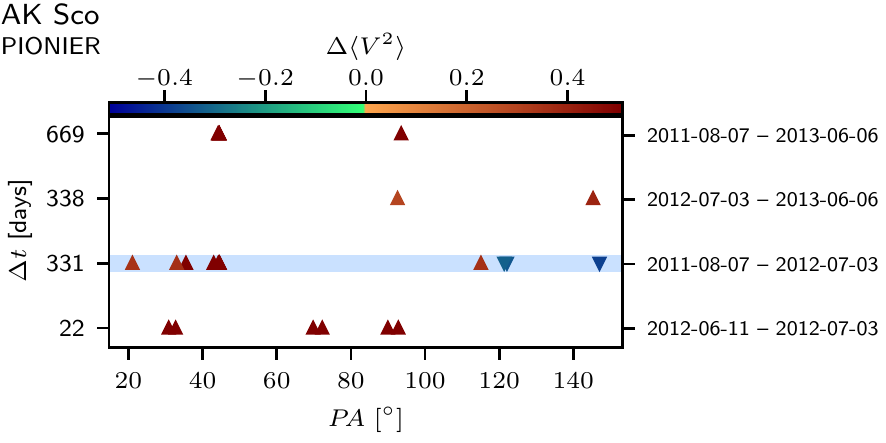}
 \includegraphics[width=\figwidthtrend]{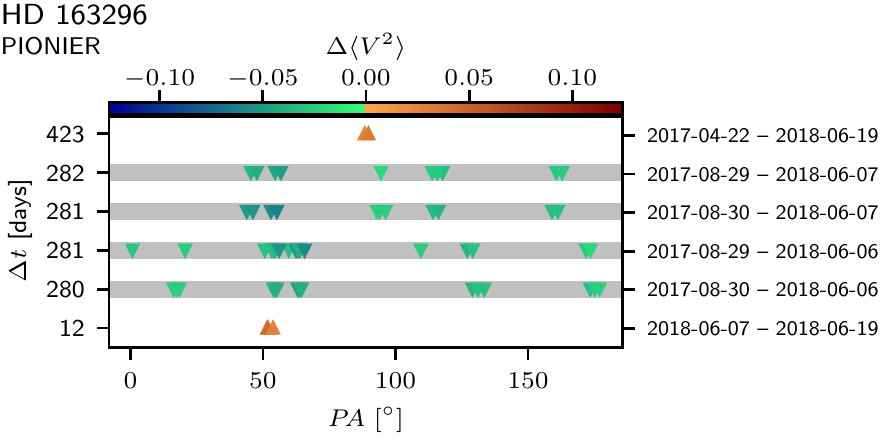}

 \includegraphics[width=\figwidthtrend]{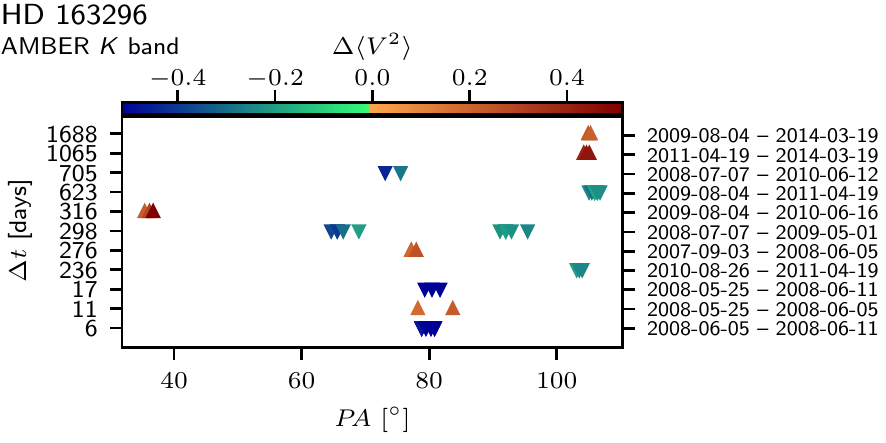}
 \includegraphics[width=\figwidthtrend]{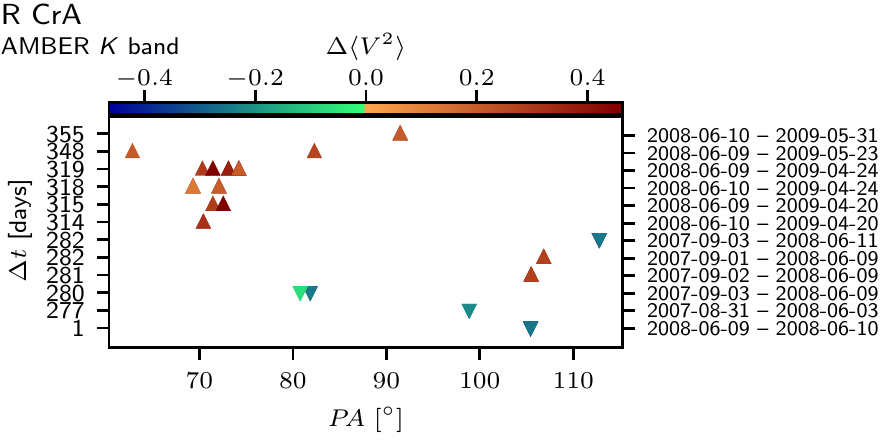}
\caption{Overview of the classification of the variations found for the different objects (see title of each plot) based on PIONIER and AMBER observations (see subtitle of each plot). For the different pairs of nights (right y-axis) with a time delay $\Delta\,t$ (left y-axis) we plot the mean position angles (x-axis) indicating the difference in the squared visibilities with the color. To emphasize the directions of the variations, we use triangles as markers (increase: tip upwards, decrease: tip downwards). The background color indicates whether the variations can be classified (supposedly symmetric: gray, asymmetric: blue)}\label{fig:trendplot}
\end{figure*}

To classify the variability of the seven objects identified as variable, we investigate the variations of the visibilities as a function of the position angle for different pairs of observation epochs. For this purpose, we consider the visibility differences that we have identified as being caused by temporal variations. First, these are pairs of visibilities measured using equal baselines differing by at least $3\sigma$ (see Sect.\ \ref{sec:dircomp}). In addition, we include those pairs obtained with deviating baselines whose visibility differences were probably caused by temporal variability ($p>0.9$, see Sect.\ \ref{sec:extComparison}). In Fig.\ \ref{fig:trendplot} we show these visibility differences together with the respective position angles and pair of nights in which the observations were obtained, thereby restricting ourselves to the pairs of nights for which variations were found for at least two position angles. It should be noted that the position angles can be very similar, such that they do not appear resolved in the plots (e.g., some night pairs with AMBER observations of R CrA).

Based on the plots shown in Fig.\ \ref{fig:trendplot}, we learn the following. If we have a pair of nights for which we find both increasing and decreasing squared visibilities, we can identify the variations as asymmetric (marked with a blue background color). If we find variations in one direction only, the conclusion regarding the symmetry of the variations becomes more reliable the better the coverage of the position angles is. In the following we classify all variations as "supposedly symmetric" (marked with a gray background color) for which the respective position angles cover a range of at least 90 degrees. The results of this classification are summarized in Table \ref{tab:sample}.

\textbf{HD 50138:} In the case of the PIONIER observations, only one pair of squared visibilities shows a significant difference (see Sect.\ \ref{sec:dircomp}), which is why the analysis regarding the symmetry is not possible. The comparison of AMBER observations from 9 January 2008 and 1 March 2009 ($\Delta t = \SI{417}{\day}$) gives both decreasing and increasing visibilities, showing an asymmetric variation. The sparse coverage of position angles for the pair 1 March 2009 -- 16 November 2009 does not allow conclusions regarding the symmetry of the variations.\\
\textbf{DX Cha:} Comparing PIONIER observations we find asymmetric variations on a large range of timescales ranging from \SIrange{3}{ 732}{\day}. As DX Cha is a close binary \citep[\SI{2}{mas},][]{2013MNRAS.430.1839G} with a period of 20\,days \citep{2004A&A...427..907B}, these are expected. However, comparing other pairs of nights we also find supposedly symmetrical variations on timescales of \SI{\sim 50}{\day} and \SI{324}{\day} (increase: 19 February 2013 -- 11 April 2013, 18 February 2013 -- 11 April 2013; decrease: 11 April 2013 -- 1 March 2014). AMBER observations on 27 May 2008 and 19 May 2009 ($\Delta t = \SI{357}{\day}$) also show supposedly symmetric variations. For these variations, azimuthally symmetrical variations of the brightness distribution are a possible explanation. \\
\textbf{HD 142527:}  The variations found in the $H$ band, with visibility differences up to 0.34 (see Sect.\ \ref{sec:dircomp}), are much larger than the variations we would expect based on the orbital motion of the companion ($\lesssim 0.06$, see Sect.\ \ref{sec:previous_studies}). However, we cannot constrain other potential origins, as analyzing the symmetry of the variations is only possible for one pair of nights, where we cannot find any signs of asymmetric variations. At the same time, the small range of position angles covered by the observations does not allow the classification as supposedly symmetrical either.  \\
\textbf{V856 Sco:} For both PIONIER and AMBER we find asymmetric variations (PIONIER: 25 May 2012 -- 22 June 2014, AMBER: 3 June 2008 -- 9 June 2010). In both bands, the asymmetric variations occur only on the largest timescales covered (PIONIER: \SI{819}{\day}, AMBER: \SI{736}{\day}). Additionally, for several pairs of nights we find supposedly symmetric variations on timescales of \SIrange{83}{349}{\day} (PIONIER: 14 June 2013 -- 29 May 2014, 25 March 2014 -- 22 June 2014; AMBER: 31 May 2008 -- 5 April 2009, 5 April 2009 -- 27 June 2009, 30 April 2009 -- 18 March 2010).\\
\textbf{AK Sco:} As expected, we find asymmetric variations (PIONIER: 7 August 2011 -- 3 July 2012) for the close binary AK Sco \citep[\SI{1}{mas, }][]{1989A&A...219..142A}. Three further pairs of nights show only increasing visibilities. However, the range of position angles is too small to classify those variations as supposedly symmetrical.\\
\textbf{HD 163296:} Comparing PIONIER observations from August 2017 to observations from June 2018, we find supposedly symmetric variations (29 August 2017 -- 6 June 2018, 29 August 2017 -- 7 June 2018, 30 August 2017 -- 6 June 2018, 30 August 2017 -- 7 June 2018). In the case of AMBER, we cannot find any signs of asymmetric variations. However, the small range of considered position angles does not allow the classification as supposedly symmetrical.\\
\textbf{R CrA:} AMBER observations of R CrA show no signs of asymmetric variations. However, the small ranges of position angles do not allow us to classify the variations as supposedly symmetrical.

For the classification of the variations we have limited ourselves to the temporal changes of the squared visibilities. In the case of asymmetric variations, however, the closure phases are of particular interest. Asymmetric variations require an asymmetric structure of the brightness distribution, which is already apparent in the closure phases by significant deviations from zero. This is the case for all objects showing variations we classify as asymmetric. In contrast, the presence of asymmetries in the brightness distribution is not a clear indication of asymmetric variations. For example, with PIONIER for HD 163296, closure phases significantly differing from zero by up to \SI{\sim35}{\degree} were measured. However, we have found indications of symmetrical variations only. Nevertheless, for further investigation of the origin of the asymmetric variations, for example, based on advanced modeling, the closure phases should be taken into account as they contain the information about the axial symmetry of the brightness distribution.

In summary, in addition to the two close binaries DX Cha and AK Sco we find asymmetric variations for HD 50138 and V856 Sco caused by the motion of asymmetric structures in the brightness distributions. Interferometric observations in the $K$ band with sufficient spatial resolution to perform image reconstruction can help to identify the origin of the asymmetry responsible for the variations found for HD 50138. Subsequently, we can investigate whether the variations we identify in the $K$ band are caused by the same mechanism as those found by \citet{2016A&A...591A..82K} in $H$ band observations obtained with PIONIER. Investigating the origin of the asymmetric brightness distribution causing the variations of V856 Sco in the $H$ and $K$ bands also requires taking snapshots with adequate UV coverage. In addition, observations using only a few equal baselines but with good temporal coverage could be helpful to determine the timescales and periodicity of the variations and thus to constrain the origin.

%================================================================================
%================================================================================
\section{Conclusion}\label{sec:concl}
%================================================================================
The ESO archive is a treasure chest containing unprecedented high-resolution multi-epoch observations of the potential planet-forming regions in protoplanetary disks. Our aim was to investigate whether the available data obtained with PIONIER, AMBER, and MIDI allow conclusions on the temporal variations of the brightness distributions of pre-main sequence stars to be reached.
For this purpose, we used different approaches for the analysis of multi-epoch interferometric data obtained for a sample of 68 pre-main sequence stars.
 
The quantities measured with an interferometer depend on the observed brightness distribution and on the length and orientation of the baseline. A straightforward approach to analyze the data is thus the direct comparison of multi-epoch data from observations obtained for equal baselines. In this comparison, deviations of the measured visibilities directly imply a temporal variation in the brightness distribution. Multi-epoch PIONIER and/or AMBER observations obtained for equal baselines ($\Delta BL < 0.1\%$, $\Delta PA < 0.2^\circ$) are available for nine pre-main sequence stars in our sample. HD 37806 and TW Hya, for each of which one pair of visibilities with equal baselines is available, show no significant variations. For CPD-36 6759 we find variations up to $2.8\sigma,$ which we do not consider as a clear indication of temporal variability. However, significant deviations ($\geq 3\sigma$) in the visibilities  measured for HD 50138, DX Cha, HD 142527, V856 Sco, HD 163296, and R CrA, with differences up to $\Delta V^2 = 0.5$, clearly indicate temporal variations in their brightness distribution. 

In order to be also able to investigate variability based on multi-epoch data obtained with different baselines, we elaborated on three further approaches and evaluated their reliability on the basis of synthetic data with a known time dependency of the brightness distribution. 
A good overview of the data can be obtained by plotting the squared visibilities against the baseline length and representing the baseline orientation with the orientation of the bar marking each data point, as well as the date of observation with the color. However, as complex brightness distributions result in significant differences in visibilities obtained with apparently similar baselines, such a qualitative analysis of the data can result in false-positive detections of variability. 

In the next step, we estimated the impact of baseline variations. This way we were able to extend the direct comparison of visibilities to measurements with moderately different baselines (\SI{<10}{\percent}). This allowed us to analyze another 12 objects in addition to the 9 already investigated in the direct comparison considering only observations with equal baselines. Of these, 1 additional object, AK Sco, shows indications of variability.

The comparison with previous variability studies has shown that signs of temporal variability have already been found for six of the seven objects identified as variable (HD 50138, DX Cha, V856 Sco, AK Sco, HD 163296, and R CrA). For four of the objects (DX Cha, HD 142527, AK Sco, and R CrA), variations in the brightness distribution are also expected due to the presence of stellar companions. The analysis of the symmetry of the variations has shown that, in addition to the binaries, two other objects (HD 50138 and V856 Sco) show indications of variations caused by the motion of asymmetries in the brightness distribution.

Since MIDI was capable of measuring visibilities for only one baseline per observation, the UV coverage of the available multi-epoch data is not sufficient to detect temporal variability for the objects in this sample based on the presented methods. The consequence is that if one wants to combine these data with future observations with the VLTI instrument MATISSE \citep[Multi AperTure mid-Infrared SpectroScopic Experiment;][]{2014Msngr.157....5L}

 to investigate variability on timescales of up to $\sim 20$ years, these should be scheduled in such a way that the UV coverage includes baselines equal to those used for the MIDI observations, allowing a direct comparison of the visibilities without introducing ambiguities into the analysis due to baseline deviations.

\begin{acknowledgements}
J.K. gratefully acknowledges support from the DFG grants WO 857/13-1, WO 857/15-1, and WO 857/17-1. R.B. acknowledges support from the DFG grant WO 857/18-1.
This research has made use of the services of the ESO Science Archive Facility.
This research has made use of the Jean-Marie Mariotti Center \texttt{OiDB} service \footnote{Available at \url{http://oidb.jmmc.fr}}.
This research has made use of the  \texttt{AMBER data reduction package} of the Jean-Marie Mariotti Center\footnote{Available at \url{http://www.jmmc.fr/amberdrs}}.
\end{acknowledgements}

  \bibliographystyle{aa} % style aa.bst
  \bibliography{bibtex} % your references Yourfile.bib
  
\begin{appendix} 
\section{Log of the observations}

{
\tablecaption{Log of the PIONIER observations.\label{tab:logPIO}} 
\tablehead{\toprule\midrule                 % inserts double horizontal lines
Object & Obs. date & Prog. ID & Config. \\    % table heading 
\midrule}
\tablefirsthead{\toprule\midrule                 % inserts double horizontal lines
Object & Obs. date & Prog. ID & Config. \\    % table heading 
\midrule}
\begin{supertabular}{cccH}        % centered columns (4 columns)

UX Ori & 2012-12-21T06:38 & 190.C-0963(C) & A1-B2-C1-D0\\ 
 & 2012-12-22T05:37 & 190.C-0963(C) & A1-B2-C1-D0\\ 
 & 2013-01-29T01:27 & 190.C-0963(A) & G1-J3-K0\\ 
 & 2013-02-19T03:01 & 190.C-0963(B) & H0-I1\\ 
 & 2013-02-20T01:13 & 190.C-0963(B) & D0-G1-H0-I1\\ 
 & 2017-12-25T06:22 & 0100.C-0278(J) & D0-G2-J3-K0\\ 
 & 2017-12-25T07:18 & 0100.C-0278(J) & D0-G2-J3-K0\\ 
\midrule 
CO Ori & 2012-12-22T01:59 & 190.C-0963(C) & A1-C1-D0\\
\bottomrule
\end{supertabular}
}
\vspace{2cm}
{
\tablecaption{Log of the AMBER observations.\label{tab:logAMB}} 
\tablehead{\toprule\midrule                 % inserts double horizontal lines
Object & Obs. date & Prog. ID & Config. \\    % table heading 
\midrule}
\tablefirsthead{\toprule\midrule                 % inserts double horizontal lines
Object & Obs. date & Prog. ID & Config. \\    % table heading 
\midrule}
\begin{supertabular}{cccH}        % centered columns (4 columns)

RY Tau & 2008-12-14T04:08 & 082.C-0893(A) & U1-U2-U4\\ 
 & 2015-12-23T02:59 & 096.C-0293(A) & U1-U3-U4\\ 
 & 2015-12-23T03:17 & 096.C-0293(A) & U1-U3-U4\\ 
\midrule 
AB Aur & 2010-12-18T04:10 & 086.C-0684(A) & U2-U3-U4\\ 
 & 2010-12-18T04:27 & 086.C-0684(A) & U2-U3-U4\\ 
 & 2012-12-02T07:06 & 090.C-0415(A) & A1-J3-K0\\ 
 & 2012-12-07T05:41 & 090.C-0070(B) & D0-H0-I1\\ 
 & 2015-01-01T03:26 & 094.C-0721(A) & U1-U3-U4\\
\bottomrule
\end{supertabular}
}
\vspace{2cm}
{
\tablecaption{Log of the MIDI observations.\label{tab:logMIDI}} 
\tablehead{\toprule\midrule                 % inserts double horizontal lines
Object & Obs. date & Prog. ID & Config. \\    % table heading 
\midrule}
\tablefirsthead{\toprule\midrule                 % inserts double horizontal lines
Object & Obs. date & Prog. ID & Config. \\    % table heading 
\midrule}
\begin{supertabular}{cccH}        % centered columns (4 columns)

RY Tau & 2004-11-01T04:43 & 074.C-0209(A) & U4-U2\\ 
 & 2004-11-01T04:57 & 074.C-0209(A) & U4-U2\\ 
 & 2004-11-04T07:51 & 074.C-0209(A) & U3-U4\\ 
 & 2004-12-28T01:47 & 274.C-5032(A) & U2-U3\\ 
 & 2013-12-30T03:27 & 092.C-0086(D) & K0-J3\\ 
 & 2013-12-30T03:36 & 092.C-0086(D) & K0-J3\\ 
 & 2014-02-07T01:06 & 092.C-0086(C) & D0-H0\\ 
\midrule 
T Tau S & 2004-10-31T07:33 & 074.C-0209(A) & U3-U2\\
\bottomrule
\end{supertabular}
}

\section{Simple model fitting approach \label{sec:fitting}}
A commonly used approach to analyze interferometric observations is to fit a Gaussian brightness distribution to reproduce the measured squared visibilities and closure phases. In the following approach, the basic idea is that instead of the measured visibilities, we compare best-fit models that individually reproduce the measured visibilities from each epoch.

In detail, we proceed as follows. We fit the squared visibilities and closure phases obtained in different epochs individually with a Gaussian brightness distribution, which has the following visibility function:
\begin{equation}\label{eq:gauss}
V(u,v) = \exp\left(\frac{-\pi^2 \mathit{FWHM}^2 }{4 \ln{2}} 
\left( \tilde{u}^2 + \tilde{v}^2 \right) \right),\\
\end{equation}
\begin{equation}
\tilde{u} = u \cos{\theta} + v \sin{\theta},\\
\end{equation}
\begin{equation}
\tilde{v} = \frac{1}{\cos{\phi}} \left( v \cos{\theta} - u \sin{\theta} \right),
\end{equation}
where $\mathit{FWHM}$ is the full width at half maximum, $\phi$ is the inclination, and $\theta$ the position angle. Due to the point symmetry of this function, the interferometric phase is zero. To determine the goodness of fit, we calculate the reduced chi-squared for $N$ measured squared visibilities and phases
\begin{equation}
X_\mathrm{red}^2 = \frac{1}{\mathit{f}}\sum_{i=1}^{N} \frac{\left(D_{i} - M_{i}\right)^2}{\sigma_{i}^2},
\end{equation}
where $\mathit{f}$ is the degree of freedom, $D$ the observed data (squared visibilities and phases), $M$ the squared visibilities and phases calculated from the model, and $\sigma$ the measurement uncertainties.

In order to compare the brightness distribution of two different epochs $j$ and $k$, in addition to the goodness of fit to the data of the corresponding epoch, we calculate whether the data of the other epochs are reproduced by the same model. Thus, we calculate four quantities: 
\begin{itemize}
 \item[1.)] A reduced chi-squared $X_\mathrm{red}^2(D_j, M_j)$ giving the goodness of fit of the best-fit model $M_j$ to the data obtained in the corresponding epoch $j$. 
 \item[2.)] The equivalent reduced chi-squared $X_\mathrm{red}^2(D_k, M_k)$ comparing the data $D_k$ from epoch $k$ to the model $M_k$. 
 \item[3.)] A reduced chi-squared $X_\mathrm{red}^2(D_j, M_k)$ indicating how well the data from epoch $j$ are reproduced by the model $M_k$. 
 \item[4.)] The equivalent reduced chi-squared $X_\mathrm{red}^2(D_k, M_j)$ comparing the data from epoch $k$ to the best-fit model $M_j$.
\end{itemize}
In the case of significant temporal variations, we expect the models $M_j$ and $M_k$ to reproduce their corresponding data $D_j$ and $D_k$ ($X_\mathrm{red}^2<1.5$), whereas the reduced chi-squared is large when comparing the data to the best-fit model found for the other epoch ($X_\mathrm{red}^2>5$). 

For a meaningful comparison of the multi-epoch observations, the following prerequisites for the UV coverage must be fulfilled. First, the interferometric observations obtained within one epoch must have a sufficient UV coverage to ensure that the best-fit model reproduces the observed brightness distribution. To improve the UV coverage, we assume that the brightness distributions do not change significantly within one night and therefore bundle all observations obtained within one night into one epoch. However, the UV coverage is still sparse, thus the comparisons must be limited to data sets for which the UV coverage is comparable. Therefore, we only compare data from epochs where the overlap of the baseline ranges is at least $2/3$ of the total range of baseline lengths covered in both epochs.

\begin{figure}
\includegraphics[]{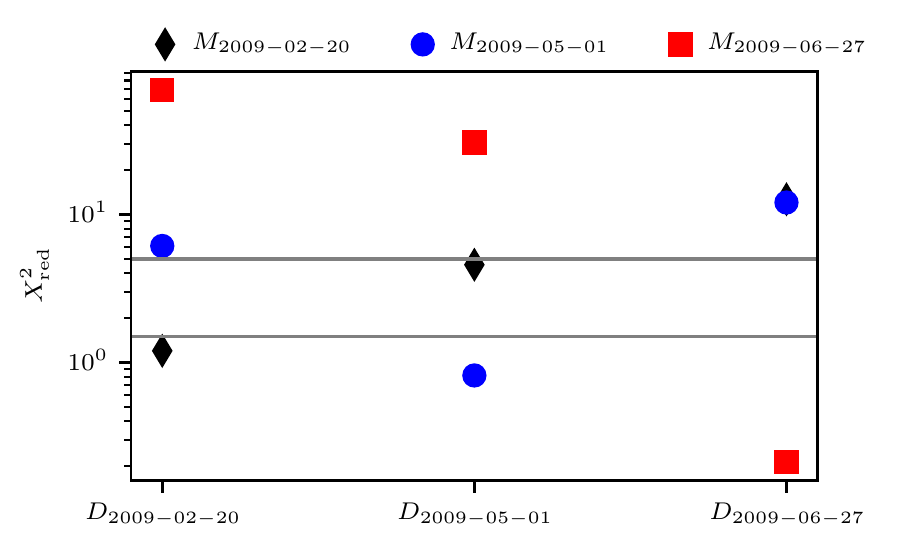}
\caption{Illustration of the analysis approach based on Gaussian model fitting using the example of interferometric observations obtained with AMBER in 20 February 2009, 1 May 2009, and 27 June 2009 for V856 Sco (see Sect.\ \ref{sec:fitting} for details). We plot
$X_\mathrm{red}^2$ to compare the Gaussian best-fit models $M_\mathrm{2009-02-20}$ (black diamonds), $M_\mathrm{2009-05-01}$ (blue dots), and $M_\mathrm{2009-06-27}$ (red squares) to the interferometric data (visibilities and phases; $D_\mathrm{2009-02-20}$, $D_\mathrm{2009-05-01}$, and $D_\mathrm{2009-06-27}$) from the different epochs. The upper and lower $X_\mathrm{red}^2$-limits indicating temporal variability when comparing data with best-fit models of the same and different epochs are plotted as gray solid lines.
\label{fig:XredV856Sco}}
\end{figure}

We illustrate this approach using the example of interferometric observations obtained with AMBER for V856 Sco. 
For V856 Sco, we find best-fit models with $X_\mathrm{red}^2<1.5$ for three epochs (20 February 2009, 1 May 2009, 27 June 2009). The baseline ranges covered on those nights overlap by at least $2/3$ allowing the comparison of the best-fit models and interferometric data of all three nights, which is illustrated in Fig.\ \ref{fig:XredV856Sco}.
The best-fit models reproduce the corresponding data from 20 February 2009, 1 May 2009, and 27 June 2009 with reduced chi-squared values of $1.2$, $0.8$, and $0.2$, respectively. Comparing the first two epochs, we find that the best-fit models $M_\mathrm{2009-02-20}$ and $M_\mathrm{2009-05-01}$ do not reproduce the interferometric data from the other respective epoch $D_\mathrm{2009-05-01}$ and $D_\mathrm{2009-02-20}$ with $X_\mathrm{red}^2(D_\mathrm{2009-05-01}, M_\mathrm{2009-02-20}) = 4.6$ and $X_\mathrm{red}^2(D_\mathrm{2009-02-20}, M_\mathrm{2009-05-01}) = 6.1$. Since the former does not meet our criterion ($X_\mathrm{red}^2>5$), we do not conclude that the variation is significant. However, when analogously comparing the second and third epoch (1 May 2009, 27 June 2009), we find significant differences between the best-fit models $M_\mathrm{2009-05-01}$ and $M_\mathrm{2009-06-27}$ and the interferometric data $D_\mathrm{2009-06-27}$ and $D_\mathrm{2009-05-01}$ from the other respective epoch ($X_\mathrm{red}^2(D_\mathrm{2009-06-27}, M_\mathrm{2009-05-01}) = 12$, $X_\mathrm{red}^2(D_\mathrm{2009-05-01}, M_\mathrm{2009-06-27}) = 30.6$). This indicates a significant difference in the best-fit models $M_\mathrm{2009-05-01}$ and $M_\mathrm{2009-06-27}$. Together with the fact that the best-fit models reproduce the interferometric data of their respective epochs ($X_\mathrm{red}^2<1.5$), this indicates that the brightness distribution of V856 Sco varied between 1 May 2009 and 27 June 2009.

\subsubsection{Evaluation with synthetic observations}
In order to evaluate the reliability of this analysis strategy, we first test whether the comparison of Gaussian best-fit models is in principle capable of identifying temporal variability based on  synthetic data.
Therefore, we calculate synthetic observations based on the variable models $M_1(t)$ and $M_2(t)$ using the temporal and spatial coverage from the real observations of all objects with available multi-epoch PIONIER data. However, only a small fraction of the synthetic data can be reproduced with a Gaussian model ($X_\mathrm{red}^2<1.5$). In particular, the presented analysis can only be applied to the synthetic data based on the model $M_1(t)$ using the spatial and temporal coverage of the observations performed for three of the objects (HD 142527, HD 144432, and HD 163296). Synthetic data based on the model $M_2(t)$ cannot be reproduced for any of the UV coverages of the considered PIONIER data.

In the next step, we apply the analysis method to synthetic data based on the time-independent models $M_{1,\mathrm{stat}}$ and $M_{2,\mathrm{stat}}$. We get a false-positive result for one of the three objects with evaluable data for $M_{1,\mathrm{stat}}$. This is due to the fact that the brightness distribution of the synthetic model is not well represented by a Gaussian distribution. However, the sparse UV coverage within individual nights partially allows the reproduction of the squared visibilities with Gaussian models. These do not represent the brightness distribution on all spatial frequencies and therefore cannot reproduce data from observations with different UV coverage.

We conclude that investigating the temporal variability of the brightness distribution of pre-main sequence stars based on Gaussian models fitted to the multi-epoch interferometric observations is impractical. Even if the Gaussian models reproduce the observations for a given UV coverage, this does not mean that the underlying brightness distribution is reproduced by the model. When comparing observations with different UV coverages this leads to false-positive conclusions on the temporal variability.

\clearpage
\onecolumn
\section{Additional table}
\setcounter{table}{0}

\begin{longtable}{lccccccccc}
\caption[]{\label{tab:sample}Overview of the sample. The separations of known binaries are listed in the fifth column, whereas for multiple systems we only give the distance to the companion nearest to the observed object. Separations smaller than the field of view of the VLTI \citep[\SI{\sim 0.16}{\arcsec},][]{2016SPIE.9907E..3BH} are indicated in bold face. The flag in the sixth column indicates the availability of multi-epoch observations obtained with PIONIER (P), AMBER (A), and MIDI (M). The seventh column indicates whether evidence for temporal variability is found based on the two approaches described in Sect.\ \ref{sec:dircomp} (A1, direct comparison of interferometric observations obtained using equal baselines) and Sect.\ \ref{sec:extComparison} (A3, comparison of interferometric observations obtained with different baselines). In the last column, the variable objects are classified according to the symmetry of the found variations (s: supposedly symmetric, a: asymmetric, -: unclassified; compare Sect.\ \ref{sec:trend})} \\
\hline\hline
Name & RA (J2000) & Dec (J2000) & d & separation & \multicolumn{3}{c}{Multi-epoch} & Variability & Classification \\
           &   (h ms s) &(${}^\circ\ {}^\prime\ {}^{\prime\prime}$) & (pc) & (${}^{\prime\prime}$)& \multicolumn{3}{c}{data} & & \\
\midrule 
\endfirsthead
\caption{continued.}\\
\hline\hline
Name & RA (J2000) & Dec (J2000) & d & separation & \multicolumn{3}{c}{Multi-epoch} & Variability & Classification \\
           &   (h ms s) &(${}^\circ\ {}^\prime\ {}^{\prime\prime}$) & (pc) & (${}^{\prime\prime}$)& \multicolumn{3}{c}{data} & & \\
\midrule 
\endhead
\hline
\endfoot
RY Tau & 04 21 57.4 & +28 26 35.5&$444^{+55}_{-44}\ {}^{(1)}$ & & - & A & M \\ 
T Tau S & 04 21 59.4 & +19 32 05.9&$144\pm2\ {}^{(1)}$ & $\mathbf{0.12}\ {}^{(3)}$ & - & - & M \\ 
T Tau N & 04 21 59.4 & +19 32 06.4&$144\pm2 {}^{(1)}$ & 0.7\ ${}^{(4)}$ & - & - & M \\ 
DG Tau & 04 27 04.7 & +26 06 16.0&$121\pm2\ {}^{(1)}$ & & - & - & M \\ 
Haro 6-10N & 04 29 23.7 & +24 33 00.9&$181^{+20}_{-16}\ {}^{(1)}$ & 1.2\ ${}^{(5)}$ & - & - & M \\ 
Haro 6-10S & 04 29 23.7 & +24 32 59.7&$181^{+20}_{-16}\ {}^{(1)}$ & 1.2\ ${}^{(5)}$ & - & - & M \\ 
GG Tau & 04 32 30.3 & +17 31 40.8&$2349^{+1061}_{-750}\ {}^{(1)}$ & $\mathbf{0.032}$\ ${}^{(6)}$ & - & - & M \\ 
DR Tau & 04 47 06.2 & +16 58 42.8&$195\pm2\ {}^{(1)}$ & & - & - & M \\ 
AB Aur & 04 55 45.8 & +30 33 04.3&$162^{+2}_{-1}\ {}^{(1)}$ & & - & A & M \\ 
HD 31648 & 04 58 46.3 & +29 50 36.0&$161\pm2\ {}^{(1)}$ & & - & - & M \\ 
UX Ori & 05 04 29.0 & -03 47 14.3&$322\pm5\ {}^{(1)}$ & & P & - & M &\\ 
CO Ori & 05 27 38.3 & +11 25 38.9&$399\pm7\ {}^{(1)}$ & 2\ ${}^{(7)}$ & P & - & M \\ 
GW Ori & 05 29 08.4 & +11 52 12.7&$398^{+11}_{-10}\ {}^{(1)}$ & $\mathbf{0.003}$\ ${}^{(8)}$ & P & - & M & \\ 
HD 36112 & 05 30 27.5 & +25 19 57.1&$160\pm2\ {}^{(1)}$ & & - & - & M \\ 
CQ Tau & 05 35 58.5 & +24 44 54.1&$162\pm2\ {}^{(1)}$ & & P & - & - \\ 
V1247 Ori & 05 38 05.3 & -01 15 21.7&$394\pm10\ {}^{(1)}$ & & P & - & - \\ 
HD 37806 & 05 41 02.3 & -02 43 00.7&$423^{+11}_{-10}\ {}^{(1)}$ & & P & - & - \\ 
FU Ori & 05 45 22.4 & +09 04 12.3&$411^{+9}_{-8}\ {}^{(1)}$ & 0.4\ ${}^{(9)}$ & P & - & M \\ 
V1647 Ori & 05 46 13.1 & -00 06 04.9&$451^{+54}_{-44}\ {}^{(1)}$ & & - & - & M \\ 
HD 45677 & 06 28 17.4 & -13 03 11.1&$614^{+23}_{-21}\ {}^{(1)}$ & & - & A & - \\ 
HD 259431 & 06 33 05.2 & +10 19 19.0&$711^{+25}_{-23}\ {}^{(1)}$ & & - & A & - \\ 
HD 50138 & 06 51 33.4 & -06 57 59.4&$377\pm9\ {}^{(1)}$ & & P & A & - & A1(P), A3(A) & a\\ 
Z CMa & 07 03 43.2 & -11 33 06.2&$252^{+118}_{-61}\ {}^{(1)}$ & $\mathbf{0.1}$\ ${}^{(10)}$ & - & - & M \\ 
HD 72106 & 08 29 34.9 & -38 36 21.1&$2552^{+2141}_{-1257}\ {}^{(1)}$ & 0.8\ ${}^{(11)}$ & - & - & M \\ 
HD 85567 & 09 50 28.5 & -60 58 02.9&$1002^{+30}_{-28}\ {}^{(1)}$ & & - & A & - \\ 
Hen 3-545 & 10 59 06.0 & -77 01 40.3&$186\pm1\ {}^{(1)}$ & & - & - & M \\ 
TW Hya & 11 01 51.9 & -34 42 17.0&$60\pm0\ {}^{(1)}$ & & P & - & M \\ 
DI Cha & 11 07 20.7 & -77 38 07.3&$190\pm1\ {}^{(1)}$ & 0.2\ ${}^{(12)}$& P & - & M \\ 
HP Cha & 11 08 15.1 & -77 33 53.2&$200\pm8\ {}^{(1)}$ & & - & - & M \\ 
Ass Cha T 1-23 & 11 09 53.4 & -76 34 25.7&$201\pm6\ {}^{(1)}$ & & - & - & M \\ 
WW Cha & 11 10 00.1 & -76 34 57.0&$191\pm1\ {}^{(1)}$ & & P & - & M \\ 
CV Cha & 11 12 27.7 & -76 44 22.3&$192\pm1\ {}^{(1)}$ & & P & - & M \\ 
HD 100453 & 11 33 05.6 & -54 19 28.5&$104\pm0\ {}^{(1)}$ & & - & A & - \\ 
DX Cha & 12 00 05.1 & -78 11 34.6&$108\pm1\ {}^{(1)}$ & $\mathbf{0.002}$\ ${}^{(13)}$ & P & A & M & A1(P,A), A3(P) & s, a\\ 
CPD-36 6759 & 15 15 48.4 & -37 09 16.0&$135\pm1\ {}^{(1)}$ & & P & A & M &  \\ %A1(P)
HD 139614 & 15 40 46.4 & -42 29 53.5&$134\pm1\ {}^{(1)}$ & & P & - & M \\ 
HD 141569 & 15 49 57.7 & -03 55 16.3&$110\pm1\ {}^{(1)}$ & & P & - & M \\ 
HD 142666 & 15 56 40.0 & -22 01 40.0&$148\pm1\ {}^{(1)}$ & & - & A & - \\ 
HD 142527 & 15 56 41.9 & -42 19 23.2&$157\pm1\ {}^{(1)}$ & $\mathbf{0.08}$\ ${}^{(14)}$ & P & A & M & A1(P,A) & -\\ 
RU Lup & 15 56 42.3 & -37 49 15.5&$159\pm2\ {}^{(1)}$ & & P & - & M \\ 
HD 143006 & 15 58 36.9 & -22 57 15.2&$165\pm4\ {}^{(1)}$ & & P & - & - \\ 
HD 325367 & 16 03 05.5 & -40 18 25.4&$157\pm1\ {}^{(1)}$ & & - & - & M \\ 
HD 144432 & 16 06 57.0 & -27 43 09.8&$155\pm1\ {}^{(1)}$ & 1.47\ ${}^{(15)}$ & P & - & M \\ 
V856 Sco & 16 08 34.3 & -39 06 18.3&$160\pm2\ {}^{(1)}$ & & P & A & M & A1(P,A), A3(P,A) & s, a\\ 
AS 205 N & 16 11 31.4 & -18 38 26.3&$127\pm2\ {}^{(1)}$ & 1.3\ ${}^{(16)}$ & P & - & - &\\ 
SR 4 & 16 25 56.2 & -24 20 48.2&$134\pm1\ {}^{(1)}$ & & - & - & M \\ 
Haro 1-6 & 16 26 03.0 & -24 23 36.2&$134\pm1\ {}^{(1)}$ & & - & - & M \\ 
GSS 31 & 16 26 23.4 & -24 20 59.6&$137\pm2\ {}^{(1)}$ & & - & - & M \\ 
Elia 2-24 & 16 26 24.1 & -24 16 13.5&$136\pm2\ {}^{(1)}$ & & - & - & M \\ 
SR 24A & 16 26 58.5 & -24 45 36.7&$114^{+5}_{-4}\ {}^{(1)}$ & 5.2\ ${}^{(7)}$ & P & - & - \\ 
SR 21A & 16 27 10.3 & -24 19 12.6&$138\pm1\ {}^{(1)}$ & 6.7\ ${}^{(17)}$ & P & - & M & \\ 
SR 9 & 16 27 40.3 & -24 22 04.1&$130\pm1\ {}^{(1)}$ & & P & - & - \\ 
Haro 1-16 & 16 31 33.5 & -24 27 37.3&$145\pm1\ {}^{(1)}$ & & P & - & M \\ 
V346 Nor & 16 32 32.2 & -44 55 30.7&$620\ {}^{(2)}$ & & - & - & M \\ 
HD 150193 & 16 40 17.9 & -23 53 45.2&$150\pm2\ {}^{(1)}$ & 1.1\ ${}^{(18}$ & P & A & M &\\ 
AS 209 & 16 49 15.3 & -14 22 08.6&$121\pm1\ {}^{(1)}$ & & - & - & M \\ 
AK Sco & 16 54 44.8 & -36 53 18.6&$140\pm1\ {}^{(1)}$ & $\mathbf{0.001}$\ ${}^{(19)}$& P & - & - & A3(P) & a\\ 
HD 163296 & 17 56 21.3 & -21 57 21.9&$101\pm1\ {}^{(1)}$ & & P & A & M & A1(A), A3(P,A) & s\\ 
HD 169142 & 18 24 29.8 & -29 46 49.3&$114\pm1\ {}^{(1)}$ & & P & - & M \\ 
MWC 297 & 18 27 39.5 & -03 49 52.1&$372\pm12\ {}^{(1)}$ & & - & A & - \\ 
VV Ser & 18 28 47.9 & +00 08 39.9&$415\pm8\ {}^{(1)}$ & & P & - & - \\ 
SVS20N & 18 29 57.7 & +01 14 05.7&$2669^{+2424}_{-1458}\ {}^{(1)}$ & 1.58\ ${}^{(20)}$ & - & - & M \\ 
SVS20S & 18 29 57.7 & +01 14 05.7&$2669^{+2424}_{-1458}\ {}^{(1)}$ & 0.32\ ${}^{(20)}$ & - & - & M \\ 
S CrA N & 19 01 08.6 & -36 57 19.9&$152\pm2\ {}^{(1)}$ & 1.3\ ${}^{(7)}$ & P & - & M \\ 
R CrA & 19 01 53.7 & -36 57 08.1&$95^{+7}_{-6}\ {}^{(1)}$ & $\mathbf{0.012}$\ ${}^{(21)}$ & - & A & - & A1(A), A3(A) & - \\ 
VV CrA NE & 19 03 06.7 & -37 12 49.7&$149\pm2\ {}^{(1)}$ & 1.9\ ${}^{(7)}$ & - & - & M \\ 
VV CrA SW & 19 03 06.7 & -37 12 49.7&$149\pm2\ {}^{(1)}$ & 1.9\ ${}^{(7)}$ & - & - & M \\ 
HD 179218 & 19 11 11.3 & +15 47 15.6&$264\pm3\ {}^{(1)}$ & & P & - & M \\ 
\end{longtable}
\tablebib{(1)~\citet{2018AJ....156...58B}, (2)~\citet{1994ApJ...424..793E}, (3)~\citet{2009A&A...502..623R}, (4)~\citet{1982ApJ...255L.103D}, (5)~\citet{1989ApJ...342L..39L}, (6)~\citet{2014A&A...565L...2D}, (7)~\citet{1993A&A...278...81R}, (8)~\citet{1991AJ....101.2184M}, (9)~\citet{2004ApJ...601L..83W}, (10)~\citet{1991AJ....102.2073K}, (11)~\citet{2000A&A...363..991D}, (12)~\citet{2013A&A...557A..80S}, (13)~\citet{2013MNRAS.430.1839G}, (14)~\citet{2014ApJ...781L..30C, 2012ApJ...753L..38B}, (15)~\citet{2011A&A...535L...3M}, (16)~\citet{1993AJ....106.2005G}, (17)~\citet{2003ApJ...591.1064B}, (18)~\citet{2017ApJ...838...20M}, (19)~\citet{1996ApJ...458..312J}, (20)~\citet{2007A&A...476..229D}, (21)~\citet{2019A&A...630A.132S}.}

\end{appendix} 
\end{document}